\documentclass[twocolumn]{aastex631}

\newcommand{\oii}{[O\,\textsc{ii}]}
\newcommand{\oiii}{[O\,\textsc{iii}]}

\newcommand{\ha}{H$\rm\alpha$}
\newcommand{\hb}{H$\rm\beta$}
\newcommand{\hg}{H$\rm\gamma$}
\newcommand{\hd}{H$\rm\delta$}
\newcommand{\lya}{Ly$\rm\alpha$}

\newcommand{\ciii}{C\,\textsc{iii}]}
\newcommand{\civ}{C\,\textsc{iv}}
\newcommand{\heii}{He\,\textsc{ii}}
\newcommand{\oiiiuv}{O\,\textsc{iii}]}

\newcommand{\neiii}{[Ne\,\textsc{iii}]}

\newcommand{\fescC}{$f_{esc}^{LyC}$}
\newcommand{\fescLy}{$f_{esc}^{Ly\alpha}$}

\newcommand{\lm}{$\lambda$~}
\newcommand{\lmlm}{$\lambda\lambda$~}
\newcommand{\kms}{km s$^{-1}$}

\usepackage[figuresright]{rotating}
\usepackage{tabularx}

\submitjournal{ApJ}

\shortauthors{Kumari et al.}

\begin{document}

\title{JADES: Physical properties of Ly$\alpha$ and non-Ly$\alpha$ emitters at z$\sim$4.8-9.6}

\correspondingauthor{Nimisha Kumari}
\email{kumari@stsci.edu}

\author[0000-0002-5320-2568]{Nimisha Kumari}
\affiliation{AURA for European Space Agency (ESA), ESA Office, Space Telescope Science Institute, 3700 San Matin Drive, Baltimore, MD, 21218, USA}

\author[0000-0001-8034-7802]{Renske Smit}
\affiliation{Astrophysics Research Institute, Liverpool John Moores University, Liverpool, L35 UG, UK}

\author[0000-0002-7595-121X]{Joris Witstok}
\affiliation{Kavli Institute for Cosmology, University of Cambridge, Madingley Road, Cambridge, CB3 0HA, UK}
\affiliation{Cavendish Laboratory, University of Cambridge, 19 JJ Thomson Avenue, Cambridge, CB3 0HE, UK}

\author[0000-0002-7626-6361]{Marco Sirianni}
\affiliation{European Space Agency (ESA), ESA Office, Space Telescope Science Institute, 3700 San Martin Drive, Baltimore, MD 21218, USA}

\author[0000-0002-4985-3819]{Roberto Maiolino}
\affiliation{Kavli Institute for Cosmology, University of Cambridge, Madingley Road, Cambridge, CB3 0HA, UK}
\affiliation{Cavendish Laboratory, University of Cambridge, 19 JJ Thomson Avenue, Cambridge, CB3 0HE, UK} 
\affiliation{Department of Physics and Astronomy, University College London, Gower Street, London WC1E 6BT, UK}

\author[0000-0002-8651-9879]{Andrew J.\ Bunker}
\affiliation{Department of Physics, University of Oxford, Denys Wilkinson Building, Keble Road, Oxford OX1 3RH, UK}

\author[0000-0003-0883-2226]{Rachana Bhatawdekar}
\affiliation{
European Space Agency (ESA), European Space Astronomy Centre (ESAC), Camino Bajo del Castillo s/n, 28692 Villanueva de la Cañada, Madrid, Spain}

\author[0000-0003-4109-304X]{Kristan Boyett}
\affiliation{School of Physics, University of Melbourne, Parkville 3010, VIC, Australia}
\affiliation{ARC Centre of Excellence for All Sky Astrophysics in 3 Dimensions (ASTRO 3D), Australia}

\author[0000-0002-0450-7306]{Alex J.\ Cameron}
\affiliation{Department of Physics, University of Oxford, Denys Wilkinson Building, Keble Road, Oxford OX1 3RH, UK}

\author[0000-0002-6719-380X]{Stefano Carniani}
\affiliation{Scuola Normale Superiore, Piazza dei Cavalieri 7, I-56126 Pisa, Italy}

\author[0000-0003-3458-2275]{Stephane Charlot}
\affiliation{Sorbonne Universit\'e, CNRS, UMR 7095, Institut d'Astrophysique de Paris, 98 bis bd Arago, 75014 Paris, France}

\author[0000-0002-2678-2560]{Mirko Curti}
\affiliation{European Southern Observatory, Karl-Schwarzschild-Strasse 2, 85748 Garching, Germany}

\author[0000-0002-9551-0534]{Emma Curtis-Lake}
\affiliation{Centre for Astrophysics Research, Department of Physics, Astronomy and Mathematics, University of Hertfordshire, Hatfield AL10 9AB, UK}

\author[0000-0003-2388-8172]{Francesco D'Eugenio}
\affiliation{Kavli Institute for Cosmology, University of Cambridge, Madingley Road, Cambridge, CB3 0HA, UK} 
\affiliation{Cavendish Laboratory, University of Cambridge, 19 JJ Thomson Avenue, Cambridge, CB3 0HE, UK}

\author[0000-0002-2929-3121]{Daniel J.\ Eisenstein}
\affiliation{Center for Astrophysics $|$ Harvard \& Smithsonian, 60 Garden St., Cambridge MA 02138 USA}

\author[0000-0003-4565-8239]{Kevin Hainline}
\affiliation{Steward Observatory, University of Arizona, 933 N. Cherry Avenue, Tucson, AZ 85721, USA}
 
 \author[0000-0001-7673-2257]{Zhiyuan Ji}
 \affiliation{Steward Observatory, University of Arizona, 933 N. Cherry Avenue, Tucson, AZ 85721, USA}

\author[0000-0002-0267-9024]{Gareth C.\ Jones}
\affiliation{Department of Physics, University of Oxford, Denys Wilkinson Building, Keble Road, Oxford OX1 3RH, UK}

\author[0000-0002-4271-0364]{Brant Robertson}
\affiliation{Department of Astronomy and Astrophysics, University of California, Santa Cruz, 1156 High Street, Santa Cruz, CA 95064, USA}

\author[0000-0001-5333-9970]{Aayush Saxena}
\affiliation{Department of Physics, University of Oxford, Denys Wilkinson Building, Keble Road, Oxford OX1 3RH, UK} \affiliation{Department of Physics and Astronomy, University College London, Gower Street, London WC1E 6BT, UK}

\author[0000-0001-6010-6809]{Jan Scholtz}
\affiliation{Kavli Institute for Cosmology, University of Cambridge, Madingley Road, Cambridge, CB3 0HA, UK} 
\affiliation{Cavendish Laboratory, University of Cambridge, 19 JJ Thomson Avenue, Cambridge, CB3 0HE, UK}

\author[0000-0003-4770-7516]{Charlotte Simmonds}
\affiliation{Kavli Institute for Cosmology, University of Cambridge, Madingley Road, Cambridge, CB3 0HA, UK}
\affiliation{Cavendish Laboratory, University of Cambridge, 19 JJ Thomson Avenue, Cambridge, CB3 0HE, UK}

\author[0000-0003-2919-7495]{Christina C.\ Williams}
\affiliation{NSF’s National Optical-Infrared Astronomy Research Laboratory, 950 North Cherry Avenue, Tucson, AZ 85719, USA}

\author[0000-0001-9262-9997]{Christopher N.\ A.\ Willmer}
\affiliation{Steward Observatory, University of Arizona, 933 N. Cherry Avenue, Tucson, AZ 85721, USA}
  
\begin{abstract}

We investigate the physical properties of Lyman-alpha emitters (LAEs) and non-Lyman-alpha emitters (non-LAEs) at z$\sim$4.8--9.6 via a stacking analysis of 253 JWST/NIRSpec spectra of galaxies observed as part of the JWST Advanced Deep Extragalactic Survey (JADES). We identify a sample of 42 LAEs with the equivalent width of \lya $\gtrsim$20\AA ~and a sample of 211 non-LAEs, divide each sample further via the median redshift of the LAEs (z$\sim$6.3), and create composite spectra using the low and medium resolution spectra from NIRSpec. We estimate physical quantities such as dust extinction, UV continuum slope $\beta$, electron temperatures, ionization parameter, escape fraction of Ly$\alpha$ and Lyman Continuum, and the photon production rate for each bin/stack. The existing dust-extinction laws do not appear to be valid at these epochs. The emission line ratio analyses show that active galactic nuclei might dominate all sub-samples, irrespective of \lya~ emission. LAEs show much higher \oiii/\oii~ and low \oii/\hd~ at z$\lesssim$6.3 compared to non-LAEs, but these line ratios are not sufficient to distinguish the two populations at z$>$6.3. However, the LAEs samples show large EW(\oiii4959, 5007) ($>$1000\AA) compared to the non-LAEs sample at all redshifts. \civ/\lya~ and \civ/\ciii ~for LAE population at z$\lesssim$6.3 is $\sim$a factor of 5 larger than that for LAE population at z$>$6.3. The ionizing radiation for LAEs is hard, as revealed from several diagnostics including \civ~detection, high \oiii/\oii ($>$8), and large values of $\xi^{\star}_{ion}$.   

\end{abstract}


\section{Introduction} 
\label{sec:intro}

\indent Building a coherent picture of the Universe's history is one of astronomy's longstanding goals. According to the Lambda cold dark model ($\Lambda$CDM), the Big Bang happened $\sim$ 13.8 Gyr ago \citep{Planck2020}, which was followed by the recombination era at redshift z$\approx$1100 when photons decoupled from baryons \citep{Peebles1970}, and are now observed as cosmic microwave background \citep[CMB;][]{Smoot1992}. The ensuing cosmic dark age ended at z$\approx$20--30 \citep{Tegmark1997, Bromm2009}, when the first stars and galaxies formed, resulting in Lyman continuum (LyC) photons capable of ionizing the neutral intergalactic medium (IGM) and thus instigating the process of cosmic reionization which lasted until z$\sim$ 5--6 \citep{Fan2006a, Bosman2022}. 

\indent While the pre-reionization epochs of the Universe can be observationally studied only via CMB, the reionization era can be explored via several direct and indirect probes. These methods encompass the studies of Gunn-Peterson absorption trough in quasar spectra \citep{Gunn1965}, CMB polarization and temperature anisotropy \citep{Kogut2003}, H \textsc{i} 21cm signal from cosmic reionization \citep{Furlanetto2004}, gamma-ray bursts \citep{Totani2006}, and potential sources of high energy photons causing reionization such as quasars, active galactic nuclei (AGN), and early star-forming galaxies including Ly-break galaxies, \lya~emitters and dusty-star-forming galaxies. See reviews by \citet{Fan2006r, Fan2023} and \citet{Stark2016r} for details on these observational probes of the reionization era.

\indent The seminal work of \citet{Partridge1967} suggested that \lya~emission is physically related to early galaxies. However, \lya~proved elusive for many years, and when z$>$3 galaxies were discovered and spectroscopically confirmed by \citep{Steidel1996}, it became clear that the equivalent width (EW) of \lya~was generally much lower than that predicted from simple stellar population synthesis models and Case B recombination (which predict EW$\sim$100--200\AA), probably through the resonantly-scattered nature of this line. In the past couple of decades, \lya~emitting galaxies (LAEs) have been identified well into the epoch of reionization thanks to careful selection techniques and exhaustive follow-up spectroscopic efforts, see e.g., \citet[][]{Bunker2003, Stark2011, Curtis-Lake2012, Cassata2015, Jiang2017} for z$\sim$6 LAEs, and \citet[][]{Ono2012, Treu2013, Pentericci2014, Jung2020} for z$>$7 LAEs. These observations allowed us to place constraints on the LAE properties such as \lya~equivalent widths \citep[][]{Cassata2015}, \lya~fraction \citep{Stark2010}, \lya~luminosity density and \lya~escape fraction \citep{Konno2018}. However, the information encoded within the rest-frame ultraviolet (UV) and optical spectra of reionization-era LAEs remained largely elusive (though not impossible, see e.g., \citet{Stark2017}) up until the launch of the James~Webb~Space~Telescope \citep[JWST;][]{Gardner2023}.

\indent The near-infrared spectrograph \citep[NIRSpec][]{Jakobsen2022} on JWST has not only pushed the \lya~detection in galaxies out to z=10.6 \citep[GNz11;][]{Bunker2023b}, but has also enabled the UV and optical spectroscopy of reionization era LAEs  \citep[see e.g.,][]{Jones2023, Saxena2023a, Saxena2023b, Jung2023, Tang2023, Maseda2023, Bunker2023b}.  These studies have revealed several physical properties of LAEs such as their dust content, ionization state, gas-phase metallicity, UV continuum slope, luminosity and \lya~escape fraction. However, these studies are mostly composed of small sample of LAEs, focus on specific set of properties, and none of them compare the properties of LAEs with the galaxies which have weak or no \lya~emission (non-LAEs hereafter) within a broad range of redshifts including the reionization era. This renders it difficult to draw broad conclusions about the LAE and non-LAE population in general.

\indent In this paper, we use JWST/NIRSpec spectroscopy of a sample of 253 galaxies in the redshift range of 4.8--9.6 to study the connection between the \lya~emission or absence thereof and the rest-frame UV and optical spectral properties. This is the first such study to employ statistically significant samples of LAEs and non-LAEs within different redshifts bins and present a comprehensive analysis of a vast range of properties encoded in the UV and optical spectra of the LAE and non-LAE populations.

\indent The paper is organized as follows: Section \ref{sec:obs} describes the observations, sample selection criteria, methodology to identify LAEs and non-LAEs, and the procedure to create the composite spectra on the basis of the strength of LAEs and redshift. In Section \ref{sec:results}, we estimate fluxes of emission lines, color-excess, electron temperatures, gas-phase metallicities, ionization parameters, and \lya~ escape fractions (\fescLy) from the composite spectra. We also estimate average UV continuum slopes $\beta$,  absolute UV magnitudes, and LyC escape fractions (\fescC) for each bin of galaxies. Section \ref{sec:discussion} explores correlations between the estimated physical quantities to distinguish the properties of different samples of LAEs and non-LAEs, such as their dust content and geometry, ionization conditions, escape fractions of \lya~ and LyC. Section \ref{sec:summary} summarizes our results.  

\indent \indent Throughout the paper, we assume a base-$\Lambda$CDM cosmology with H$_{0}$ = 67.4 \kms Mpc$^{-1}$, $\Omega_{m}$ = 0.315 \citep{Planck2020}, and an AB flux normalization \citep{Oke1990} The solar oxygen abundance is assumed to be 12 + log(O/H)$_{\odot}$ = 8.69 \citep{Asplund2009}.

\begin{figure*}
    \centering
    \includegraphics[width=0.33\textwidth]{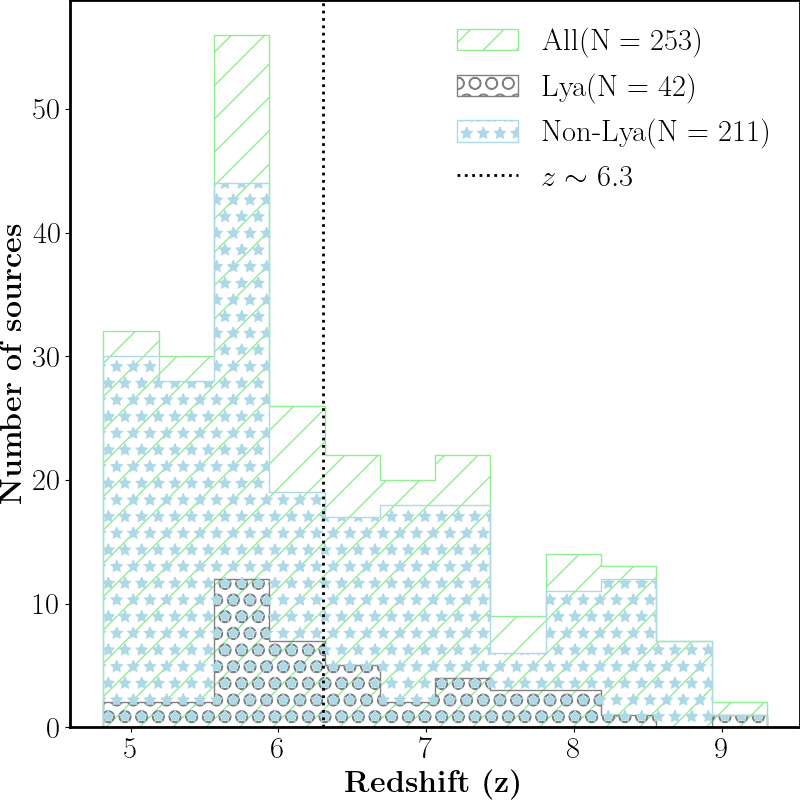}
    \includegraphics[width=0.45\textwidth]{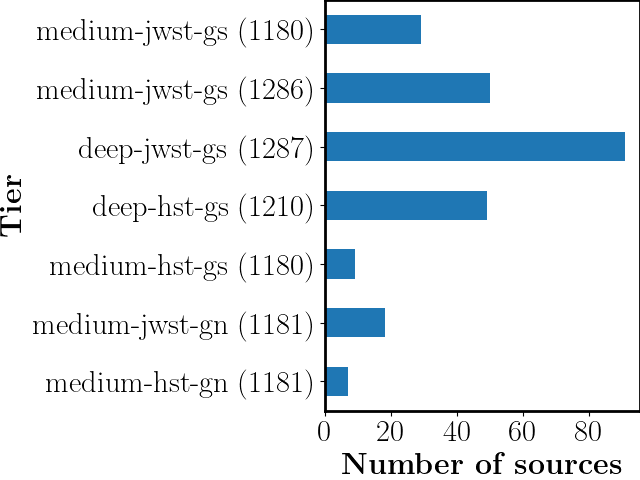}
    \caption{Left-hand panel: Redshift distribution of the final sample of 253 galaxies (light green hatched histogram), 42 \lya~emitters (grey histogram containing circles) and 211 non-\lya~emitters (light blue histogram containing stars). The vertical dotted line denotes the median redshift of the \lya~emitter sample of 42 galaxies. Right-hand panel: Tier distribution of all galaxies. Each horizontal bar denotes a Tier, and the corresponding PID of JWST observation is shown within brackets.}
    \label{fig:ztierhist}
\end{figure*}

\section{Data}
\label{sec:obs}

\subsection{The JADES survey}
\label{sec:jades}
\indent We use data observed as part of the JWST Advanced Deep Extragalactic Survey \citep[JADES;][]{Bunker2020, Rieke2020} which is based on guaranteed time observations of 770 hrs (PIDs: 1180, 1181, 1210, 1286, 1287) taken with JADES/NIRSpec micro shutter array  \citep[MSA;][]{Ferruit2022} and JWST/NIRCam. The survey focused on the Great Observatories Origins Deep Survey \citep[GOODS;][]{Giavalisco2004} deep legacy fields in both hemispheres, GOODS-South and GOODS-North, and is described in detail in \citet{Eisenstein2023}. The spectroscopic data were taken with the NIRSpec filter-grating combinations with spectral resolutions varying from $\sim$30-2700 with spectral coverage of $\sim$0.7-5 $\mu$m. Please see \citet{Bunker2023} and \citep{Deugenio2024} for the first and second data release of JADES spectroscopic data. The details of data reduction for the spectroscopic data will be provided in Carniani et al. (in prep). The NIRSpec data used in this work were taken in between 5 October 2022 and 10 January 2024.

\subsection{Sample Selection}
\label{sec:sample}
 We first determined the redshifts of all targets by inspecting their spectra by eye using the software \textsc{inzimar} developed by the JADES collaboration. The redshifts of the objects were based on either the continuum fit, a single or multiple emission lines from the spectra taken with the PRISM, medium or high-resolution gratings. More information on the redshift determination can be found in the documents provided in the first and second JADES data release. We restricted the analysis to a redshift range of 4.8--9.6, which allows us to access the rest wavelength range of $\sim$1215--5007\AA. We determined a total of 523 unique NIRSpec targets with robust redshift estimates between the imposed redshift range. We co-added the medium-resolution spectra (G140M, G235M, and G395M) of each target by first resampling them using \textsc{spectres} \citep[][]{Carnall2017} on a wavelength grid with a wavelength sampling corresponding to that of G140M and then taking a weighted average. The spectra are not necessarily contiguous throughout the wavelength range, so we further imposed a quality check that the wavelength windows for the following emission lines have finite flux values: \lya 1215, \civ \lmlm 1548,1550,  \heii\lm1640, \oiiiuv\lmlm 1660,1666, \ciii \lmlm1907,1909, \oii\lmlm 3727,3729, \neiii\lm3869, \hg4340, \oiii4363, \heii\lm4686, \hb4860, \oiii\lmlm 4959,5007. 
 The quality check on wavelength windows resulted in a total of 255 galaxies. As described in Section \ref{sec:composite}, we will use the UV continuum at 1500\AA~ for normalizing the R1000 spectra before creating the composite spectra. So, we excluded two more targets for which the UV continuum at 1500\AA~ was too weak to be detected. The final sample consists of 253 targets. Figure \ref{fig:ztierhist} (left-hand panel) shows the redshift distribution for all targets along with the sub-samples of LAEs and non-LAEs, described below. Figure \ref{fig:ztierhist}  (right-hand panel) shows the tier distribution and the associated PID of JWST observations. 

\begin{figure*}
    \centering
    \includegraphics[width=\textwidth]{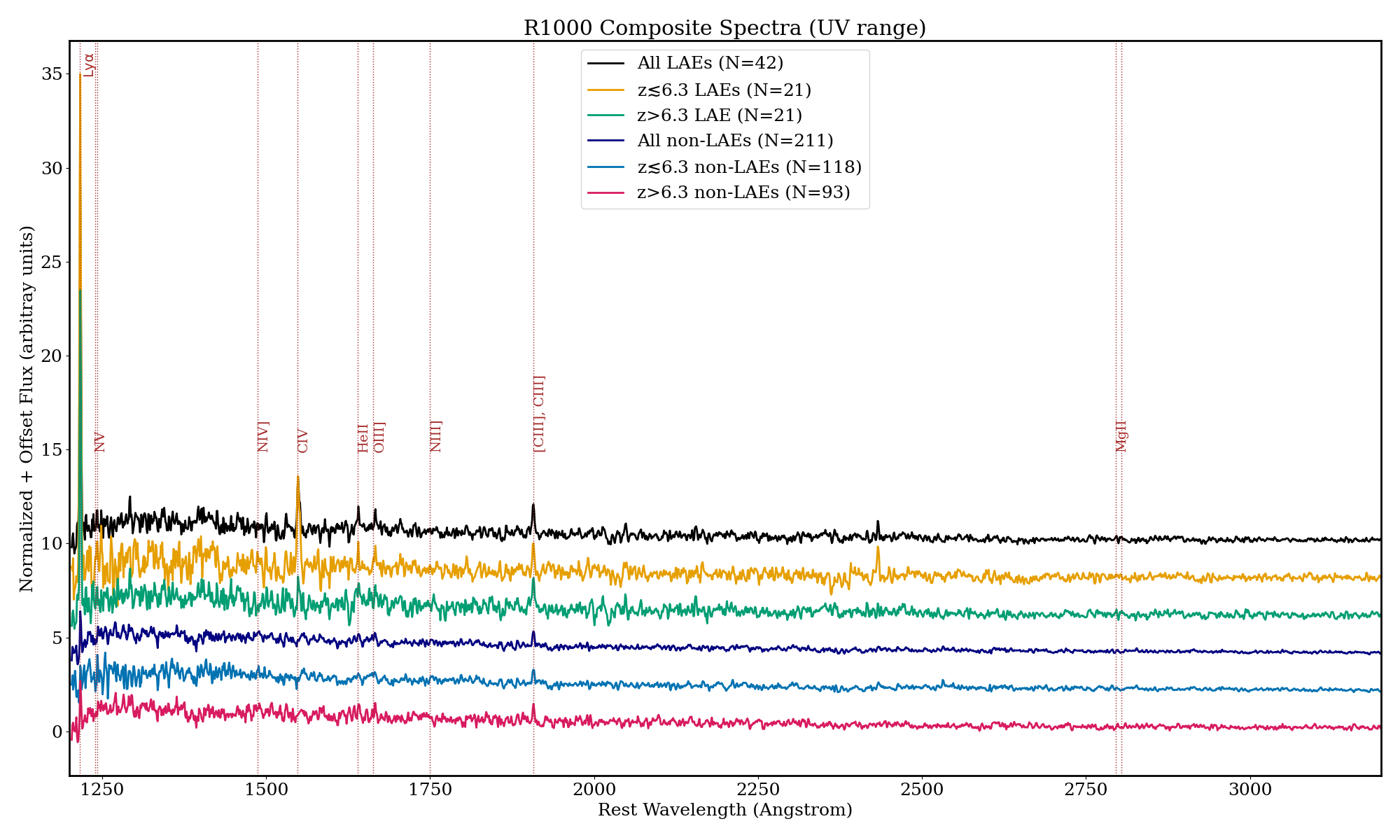}
    \includegraphics[width=\textwidth]{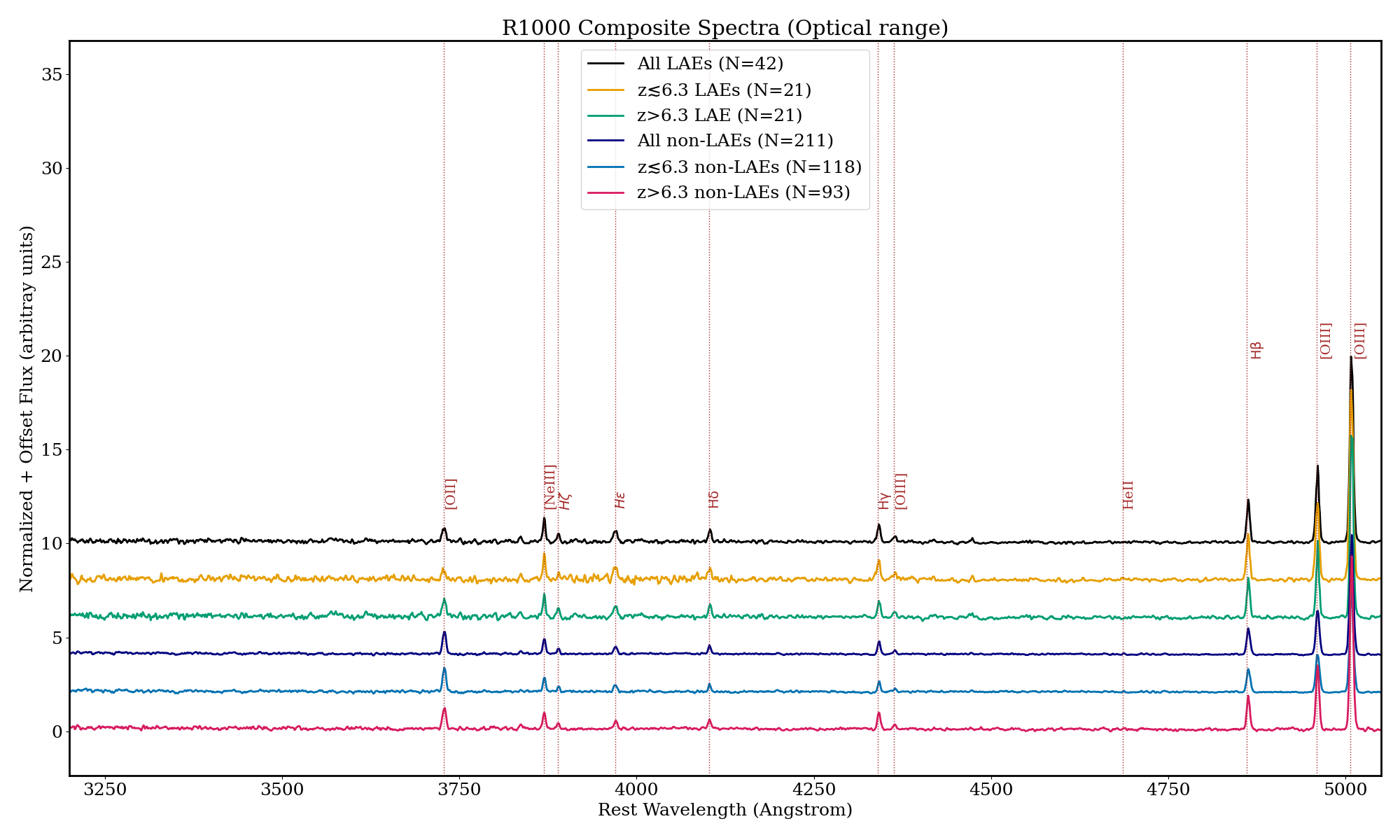}
\caption{R1000 composite spectra obtained by normalizing individual R1000 spectra with respect to the continuum at 1500\AA~from the corresponding R100 spectra. Different colors correspond to the six sub-samples of targets described in Section \ref{sec:composite}: (a) all LAEs with EW (\lya)$>$20\AA ~(black); (b) LAEs at redshifts z $\lesssim$ 6.3 (orange); (c) LAEs at redshifts $>$ 6.3 (green); (d) all non-LAEs (dark blue); (e) non-LAEs at redshift z $\lesssim$ 6.3 (light blue); (f) non-LAEs at redshifts $>$ 6.3 (magenta). The location of the UV and optical emission lines typically found in galaxies are marked in brown. Not all of the marked lines are detected in all composite spectra. The upper panel shows the UV range and the lower panel shows the optical range. In both panels, all composite spectra are offset vertically for better visibility of spectral features.}
\label{fig:stacks}
\end{figure*}

\subsection{Identifying LAEs and non-LAEs}
\label{sec:Lya}
\lya~emitters are typically defined as the sources with EW(\lya) $\gtrsim$ 20\AA~ \citep{Hayes2015}, though there are some variations across different works \citep[see e.g.,][]{Tang2023}. For estimating EW(\lya), we first estimate the \lya~flux from the R1000 spectrum for each target in our sample. However, the continua in the R1000 spectra are too faint, so we use the R100 spectra for estimating the continuum levels at 1215.67 \AA~ by fitting a power law to the spectra in the wavelength range 1300--1500 \AA\ and thereby extrapolating to 1215.67\AA. We only use the wavelength range red-ward of the \lya~ line because the R100 spectra generally show breaks at \lya~ line in the R100 spectra. We require that the \lya~ fluxes show signal-to-noise ratios (S/N) $>$ 3 and that the continua at 1215.67\AA~ are positive. The EW(\lya) is then estimated from the \lya~ fluxes from R1000 and continua from R100, which results in 42 galaxies with EW(\lya) $>$20\AA. We note that the EW(\lya) estimated in this work are in broad agreement with those found by \citet{Jones2023} and \citet{Saxena2023a} for the common targets. Figure \ref{fig:ztierhist} (left-panel, grey histogram) shows the redshift distribution of these 42 galaxies, which we refer to as the LAE sample hereafter. The rest of the 211 galaxies in the sample are referred to as the non-LAEs in this work (Figure \ref{fig:ztierhist}, light blue histogram). The median redshift for the LAE sample is z$\sim$6.28, shown by the dotted vertical line in the figure, and is used to identify sub-samples of LAEs and non-LAEs for creating composite spectra. Hence, we create 6 samples of targets: (a) 42 LAEs with EW(\lya)$>$20\AA (b) 21 LAEs lying below the median redshift z$\lesssim$6.3 (c) 21 LAEs with z$>$6.3 (d) 211 non-LAEs (e) 118 non-LAEs with  z$\lesssim$6.3 (f) 93 non-LAEs with z$>$6.3.

\subsection{Composite spectra}
\label{sec:composite}

 \indent To create composite spectra, we first de-redshift and re-sample R1000 spectra on a common wavelength grid and normalize them with the continuum at 1500\AA~ obtained from the corresponding R100 spectra. We then create 6 composite spectra corresponding to the 6 sub-samples identified in Section \ref{sec:Lya}, by taking the average of all the spectra within a sample weighted by the uncertainties on the fluxes in R1000 spectra and those on continuum determination. Different normalization schemes and weighing methods are used in the literature depending on the goals of the work and data availability \citep[see, e.g.,][]{Witstok2023, Boyett2024}. Our choice of using the continuum 1500\AA~ for normalization instead of using other parameters such as the peak flux density or the luminosity of an emission line, allowed us to include even those targets with faint emission lines by capitalizing over the availability of the R100 data for the continuum detection. Similarly, we chose to use the weighted average instead of median or mean because that allowed us to minimize the impact of noise in the data. Our experiments show that mean results in un-physical features in the stacked spectra while the median smoothens over the potential spectral features.    

 \indent Figure \ref{fig:stacks} show six R1000 composite spectra obtained by normalizing with continuum from R100, and corresponds to 6 samples shown by different colors: (a) 42 LAEs with EW(\lya)$>$20\AA (black) (b) 21 LAEs lying below the median redshift z$\lesssim$6.3 (orange) (c) 21 LAEs with z$>$6.3 (green) (d) 211 non-LAEs (dark blue) (e) 118 non-LAEs with  z$\lesssim$6.3 (light blue) (f) 93 non-LAEs with z$>$6.3 (magenta). The same color 
scheme for the samples is used in different figures in the rest of the paper. 

\indent We also created six more composite spectra solely using R100 data on a sub-sample of galaxies, for estimating EW(\oiii 4959, 5007), EW(\hb) and production rate of H-ionizing photons per unit UV luminosity, $\xi^{\star}_{ion}$. Since these composite spectra are based on a sub-sample of galaxies, we present the methodology to create the composite spectra and estimates of subsequent quantities in Appendix \ref{app:R100_composite}.

\subsection{Comparison samples from the literature}
\label{sec:literature}

\indent To compare our analysis with the literature, we selected samples such that all observable redshifts are covered, either via individual observations or composite spectra. 

\indent z$\sim$0: For the local comparison sample, we use the individual galaxies' data from \citet{Yang2017b, Izotov2020,  Flury2022}. We include five \lya~ emitters at z$\sim$0 from \citet{Izotov2020} with EW(\lya)$>$45--190\AA. The sample of \citet{Yang2017b} consists of the so-called blueberry galaxies at z$\lesssim$0.05, which are considered local analogs of high-z \lya~ emitting galaxies and are characterized by very high ionization (\oiii/\oii $\sim$ 10--60). Since \fescC~ is often associated with the properties of \lya~ emitters (EW, line profile, \fescLy), we include the galaxies from the Low-redshift Lyman Continuum Survey \citep[LzLCS;][]{Flury2022} within z$\sim$0.2--0.4. 

\indent z$\sim$1--2: We use the results from the composite spectra of \citet{Tang2019} obtained by stacking galaxies in the redshift range of 1.3--2.4 and based on the EW(\oiii\lm5007).

\indent z$>$2: For higher redshifts, we mainly rely on the results based on stacking analysis from \citet{Sanders2023, Shapley2023, Tang2023, Witstok2023} and individual galaxies' measurements from \citet{Schmidt2021} and \citet{Katz2023}.  

\section{Results}
\label{sec:results}

\subsection{Flux measurements \& Reddening correction}
\label{sec:flux}
\indent We measure the line fluxes for the recombination and collisionally excited emission lines by fitting Gaussian profiles to spectral features after subtracting a linear continuum in the spectral region of interest. We fit single Gaussian profiles to doublets such as \civ, \ciii, and \oii, because as they are unresolved in the composite spectra. However, the rest of the detected emission lines could be fit with the single Gaussian profiles. This is done for all the composite spectra described in Section \ref{sec:composite}.  

\indent For reddening correction, we use the extinction curve of \citet{Cardelli1989}, which is deemed to be appropriate for high redshift Universe \citep{Reddy2020}, and has been used in related studies \citep{Shapley2023, Witstok2023}. We estimate the color excess E(B-V) by combining the parameters of the extinction curve with the observed and intrinsic Balmer line ratio of \hg/\hb ~(=0.473), assuming a Case B recombination and electron temperature and density of 15,000 K and 100 cm$^{-3}$, respectively. We do not use the \ha/\hb~ line ratio as it is not available for high-z samples. 
The E(B-V) is then used to estimate the dust-corrected line fluxes for all emission lines. For the subsequent analysis, we do not include the large uncertainties on E(B-V) while propagating errors on dust-corrected fluxes and other different dependent quantities, as they result in unrealistic uncertainties due to the non-Gaussian nature of uncertainty distribution. The random measurement uncertainty of observed flux is propagated to estimate uncertainties on dust-corrected line fluxes and other dependent quantities throughout the work.

\begin{table*}
\centering
\caption{Properties of the composite spectra from different samples. The line ratios are obtained by using the dust-corrected fluxes from composite spectra.}
\label{tab:properties_composite}
\begin{tabular}{lcccccc}
\hline
\hline
& \multicolumn{3}{c}{LAEs} & \multicolumn{3}{c}{non-LAEs}\\
Properties & All & z$\lesssim$6.3 & z$>$6.3 & All & z$\lesssim$6.3 & z$>$6.3 \\
\hline
N$^a$ & 42 & 21 & 21 & 211 & 118 & 93 \\
& \multicolumn{6}{c}{Derived from R1000 Composite Spectra}\\
\\
RO3 & $0.034 \pm 0.002 $ & $0.033 \pm 0.004 $ & $0.033 \pm 0.003 $ & $0.029 \pm 0.002 $ & $0.033 \pm 0.003 $ & $0.023 \pm 0.002 $ \\
R23 & $6.81 \pm 0.21 $ & $6.44 \pm 0.24 $ & $7.46 \pm 0.26 $ & $7.40 \pm 0.21 $ & $7.58 \pm 0.24 $ & $7.97 \pm 0.20 $ \\
O3 & $4.66 \pm 0.15 $ & $4.61 \pm 0.17 $ & $5.02 \pm 0.18 $ & $4.63 \pm 0.13 $ & $4.25 \pm 0.14 $ & $5.39 \pm 0.14 $ \\
O3Hg & $0.34 \pm 0.03 $ & $0.32 \pm 0.04 $ & $0.35 \pm 0.03 $ & $0.28 \pm 0.02 $ & $0.30 \pm 0.03 $ & $0.26 \pm 0.02 $ \\
O2Hb & $0.56 \pm 0.04 $ & $0.27 \pm 0.03 $ & $0.73 \pm 0.06 $ & $1.20 \pm 0.04 $ & $1.87 \pm 0.08 $ & $0.76 \pm 0.03 $ \\
O32 & $8.32 \pm 0.58 $ & $17.17 \pm 2.12 $ & $6.84 \pm 0.50 $ & $3.86 \pm 0.10 $ & $2.27 \pm 0.07 $ & $7.11 \pm 0.23 $ \\
Ne3O2 & $1.11 \pm 0.09 $ & $1.51 \pm 0.21 $ & $0.99 \pm 0.09 $ & $0.54 \pm 0.02 $ & $0.45 \pm 0.02 $ & $0.67 \pm 0.03 $ \\
Ne3Hd$^b$ & $2.36 \pm 0.12 $ & $1.55 \pm 0.12 $ & $2.76 \pm 0.17 $ & $2.47 \pm 0.09 $ & $3.21 \pm 0.15 $ & $1.92 \pm 0.08 $ \\
O2Hd$^b$ & $2.13 \pm 0.16 $ & $1.02 \pm 0.13 $ & $2.79 \pm 0.22 $ & $4.57 \pm 0.17 $ & $7.14 \pm 0.30 $ & $2.89 \pm 0.12 $ \\
\lya/\hb & $52.12 \pm 2.66 $ & $12.64 \pm 0.53 $ & $50.33 \pm 3.20 $ & -- & -- & -- \\
\civ/\ciii$^c$ & $1.74 \pm 0.25 $ & $3.93 \pm 0.57 $ & $0.70 \pm 0.24 $ & -- & -- & -- \\
\civ/\lya$^c$ & $0.09 \pm 0.01 $ & $0.24 \pm 0.02 $ & $0.05 \pm 0.01 $ & -- & -- & -- \\
EW(\civ)$^c$ & $ 19 \pm 2$ & $ 39 \pm 3 $ & $ 9 \pm 3 $& -- & -- & -- \\
\\ 

Te([OIII]) (K) & $20100 \pm 800 $ & $19800 \pm 1600 $ & $19800 \pm 900 $ & $18100 \pm 600 $ & $19800 \pm 900 $ & $16000 \pm 600 $ \\
Te([OII]) (K) & $19500 \pm 700 $ & $19200 \pm 1300 $ & $19200 \pm 800 $ & $17800 \pm 500 $ & $19200 \pm 700 $ & $16000 \pm 500 $ \\
12+log(O/H)$^d$ & $7.45 \pm 0.04 $ & $7.44 \pm 0.07 $ & $7.50 \pm 0.04 $ & $7.58 \pm 0.03 $ & $7.51 \pm 0.03 $ & $7.73 \pm 0.04 $ \\
\\

log(U) & $-2.284 \pm 0.024 $ & $-2.032 \pm 0.043 $ & $-2.352 \pm 0.025 $ & $-2.551 \pm 0.009 $ & $-2.736 \pm 0.011 $ & $-2.339 \pm 0.011 $ \\

\\
& \multicolumn{6}{c}{Average of properties measured from each spectrum$^e$}\\
\\
$\rm M_{UV}$ & $-18.72 \pm 0.19$ & $-18.44 \pm 0.32$  & $-19.00 \pm 0.18$ & $-18.86 \pm 0.07$ & $-18.69 \pm 0.11$ & $-19.08 \pm 0.08 $\\
$\beta$ & $-2.26 \pm 0.09$ & $-2.19 \pm 0.12$ & $-2.34 \pm 0.12$ & $-2.09\pm0.05 $ & $-1.96 \pm 0.06$ & $-2.26\pm 0.07$\\
\\
\hline 

\hline

\end{tabular}
\\
Notes:
$^a$: N indicates the number of galaxies in each bin.\\
$^b$: For estimating line ratios involving \hd, we combined the line ratios with respect to \hb~(i.e., \neiii/\hb ~and \oii/\hb) and used the theoretical line ratio of \hb/\hd = 3.81 obtained using \textsc{pyneb} for T$_e$=15kK and N$_e$=100cm$^{-3}$.
$^c$: For the non-LAEs samples, we do not report the line ratios involving \civ~ as this doublet is not detected with enough S/N in the composite spectra of non-LAEs. We do not put any upper limit on \civ~values because the origin of \civ~ can be both nebular and stellar.\\
$^d$: Direct method metallicity.
$^e$: Uncertainties for the average properties is $\sigma/\sqrt{N}$, where $\sigma$ is the standard deviation of estimated values ($\beta$ and $\rm M_{UV}$) of all sources within a bin and N is the number of sources within the same bin.
The definition of different line ratios shown in the first column is as follows: RO3 = \oiii4363/\oiii5007, R23 =(\oii3727,3729 + \oiii4959,5007)/\hb, O3 = \oiii5007/\hb, O3Hg = \oiii4363/\hg, O2Hb = \oii3727,3729/\hb, O32 = \oiii5007/\oii3727,3729, Ne3O2 = \neiii3869/\oii3727,3729, Ne3Hd = \neiii3869/\hd,  O2Hd = \oii3727,3729/\hd. 
\end{table*}

\subsection{Electron temperature \& density}
\label{sec:Te}
\indent We measure electron temperature T$_e$(\oiii) characterizing the high ionization zone by using the temperature-sensitive \oiii4363/\oiii5007 emission lines ratio in the calibration given by \citet{Perez-Montero2017} which is valid in the range of 7000-25000K. The detection of auroral lines \oii 7320, 7330 are required to estimate T$_e$ of the low-ionization zone. Since this oxygen doublet lies beyond our wavelength range of analysis and hence not available, we simply use the linear relation T$_e$(\oiii)-T$_e$(\oii) from \citet{Pilyugin2009} to estimate T$_e$(\oii). 

\indent The electron density N$_e$ can be determined by the flux ratio of density-sensitive line doublets  \ciii1907, 1909 or \oii3727, 3729. However, even though both of these line doublets are detected with enough S/N ($>$3) in our composite spectra, the peaks of these lines were not well-resolved to estimate the flux of the individual line of the doublets. This prevented us from estimating the N$_e$ for our composite spectra. In the rest of the analysis, we assume an N$_e$ = 300 cm$^{-3}$, which has been observed and assumed in other studies of high-redshift studies \citep[see e.g., ][]{Sanders2024}.

\begin{figure*}
    \includegraphics[width=\textwidth]{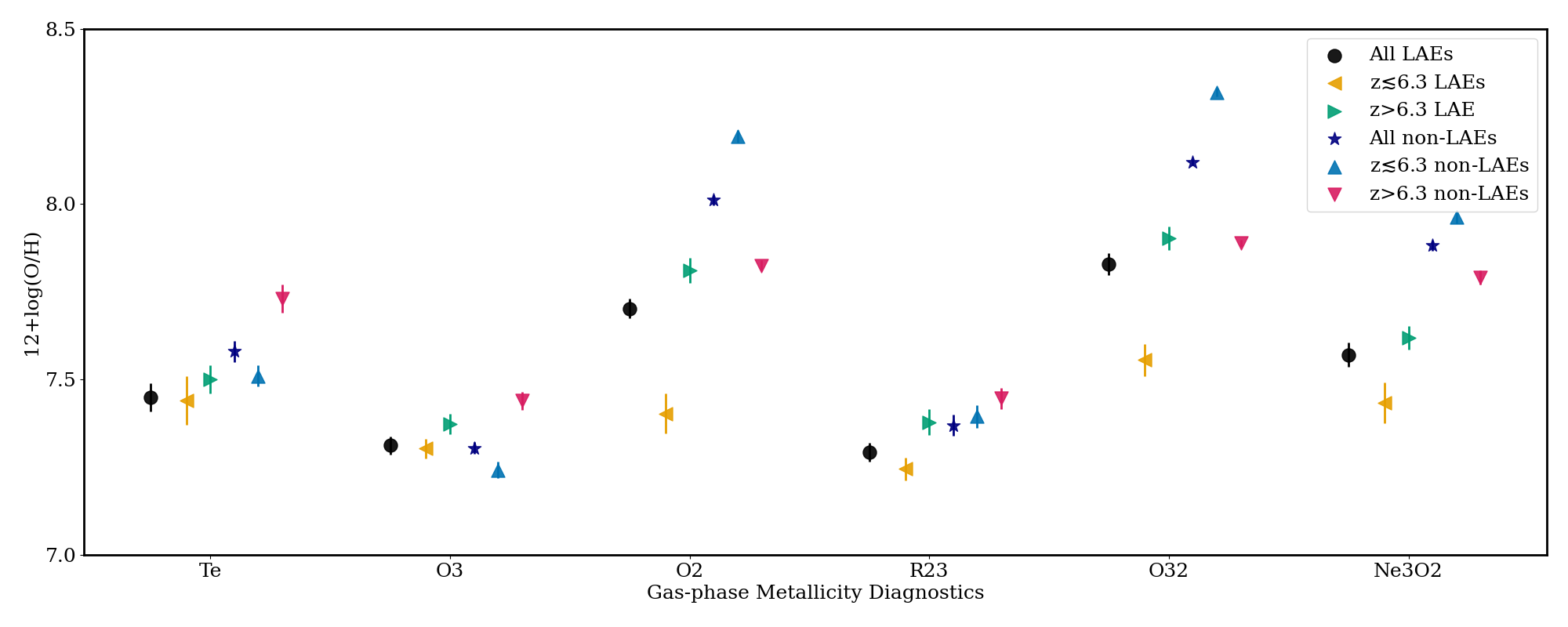}
    \caption{Comparison of gas-phase metallicities obtained via direct T$_e$-based method and the indirect method including the line ratios O3, O2, R23, O32 and Ne3O2 for all six composite spectra derived from different sub-samples: (a) all LAEs with EW (\lya)$>$20\AA ~(black circle); (b) LAEs at redshifts z $\lesssim$ 6.3 (orange left-pointing triangle); (c) LAEs at redshifts $>$ 6.3 (green right-pointing triangle); (d) all non-LAEs (dark blue stars); (e) non-LAEs at redshift z $\lesssim$ 6.3 (light blue upward-pointing triangle); (f) non-LAEs at redshifts $>$ 6.3 (magenta downward-pointing triangle). The error bars correspond to the flux measurement uncertainties propagated to the derived metallicities.}
    \label{fig:direct_vs_indirect}
\end{figure*}

\subsection{Gas-phase metallicity}
\label{sec:OH}
\indent For estimating gas-phase metallicity via direct method, we use the software \textsc{pyneb} \citep{Luridiana2015}. We measure the O$^{2+}$/H$^+$ by using the line ratio \oiii\lmlm 4959, 5007/\hb\ and the high ionization electron temperature T$_e$(\oiii). Similarly, we measure the O$^+$/H$^+$ by using the line ratio \oii\lmlm3727,3929 and the low ionization electron temperature T$_e$(\oii). We assume N$_e$=300cm$^{-3}$ in both of the above estimates. We estimate the total oxygen abundance as the sum of O$^+$/H$^+$ and O$^{2+}$/H$^+$, and estimate the gas-phase metallicity, 12+log(O/H).  

\indent We also estimate gas-phase metallicities for our composite spectra via the indirect method. In Figure \ref{fig:direct_vs_indirect}, we compare the gas-phase metallicities obtained with the direct T$_e$ method with those obtained from strong line metallicity calibrations proposed by \citet{Sanders2024}. In particular, we use the emission lines such as O3 (=\oiii/\hb), O2 (=\oii/\hb), R23 (=(\oii+\oii)/\hb), O32 (=\oiii/\oii) and Ne3O2 (=\neiii/\oii). The calibrations from \citet{Sanders2024} are devised for the galaxies from the reionization era and cosmic noon, though large scatters are found around these empirical relations. Figure \ref{fig:direct_vs_indirect} shows that the metallicities determined from O2, O32, and Ne3O2 show very high scatter which can not be explained by the intrinsic scatter on these calibrations reported by \citet{Sanders2024}. Among these line ratios showing large scatter,  the emission lines used in Ne3O2 are close in wavelength, and thus, Ne3O2 is independent of any systematics related to flux calibration. The large scatter in metallicities could be likely due to the poorly-understood ionization conditions in LAEs and non-LAEs (see Section \ref{sec:ionization}), which manifest differently in different line ratios, and care should be taken while using these calibrations. The metallicities estimated from O3 and R23 show lesser scatter.

\indent We do not try to compare these results with the calibrations for local galaxies from \citet{Curti2017} or those for the high-z analogs from \citet{Perez-Montero2021, Bian2018} and \citet{Jones2015} as our direct method metallicites lie beyond the valid range of those calibrations. In principle, one can also use the strong line calibrations from \citet{Nakajima2022} for high-z galaxies. 

\subsection{Ionization parameter}
\label{sec:logU}

\indent The ionization parameter log(U) is expected to correlate with the degree of ionization of the nebula. We estimate log(U) by using the line ratio \oiii/\oii~ in the linear relation proposed by \citet{Diaz2000} based on the single star photoionization models.  We do not use the most recent calibrations from \citet{Kewley2019} as the metallicities derived from the composite spectra of all sub-samples except one (z$>$6.3 non-LAEs) lie beyond the metallicity range of these calibrators (12+log(O/H) = 7.63--8.93). 

\subsection{\lya~ \& LyC escape fraction}
\label{sec:fesc}
\indent The \lya~escape fraction \fescLy~is defined as the ratio of the observed \lya~ (redenning-corrected) flux to the expected \lya~flux derived from the dust-corrected Balmer lines, e.g., \ha~ or \hb~(though see Section \ref{sec:dust}). We use the \hb~line flux for this purpose as the stronger \ha~line is beyond the imposed wavelength range of some of our stacks. Hence, we estimate \fescLy~using the following formulae and assuming the theoretical \lya/\hb=23.3 for a Case B recombination and electron temperature and density of 15,000 K and 100 cm$^{-3}$, respectively: \fescLy = f$_{Ly\alpha}$/(23.3 $\times$ f$_{H\beta}$), where f$_{Ly\alpha}$ and f$_{H\beta}$ are the dust-corrected line fluxes. The dust-corrected flux line ratio of \lya/\hb~ is found to be larger than the theoretical line ratio 23.3 (see Table \ref{tab:properties_composite}) (see Sections \ref{sec:dust} for possible causes). We also estimate \fescLy~ assuming no dust-extinction for all stacked spectra and discuss them wherever relevant.   

\indent We estimate the Lyman Continuum escape fraction \fescC~ by using the relation between the UV continuum slope $\beta$ and \fescC~ given by \citet{Chisholm2022}. 

\subsection{UV continuum slope $\beta$ \& absolute magnitude \rm M$_{UV}$}
\label{sec:beta_MUV}
\indent We estimate the UV continuum slope $\beta$ of all sample targets by using R100 spectra in the spectral windows from \citet{Calzetti1994} in the wavelength range of $\sim$ 1250--2580\AA.  We estimate the absolute magnitude at 1500\AA ~of all sample targets by using their R100 spectra. To do so, we first smooth the R100 spectra by a median filter of 11 pixels, then estimate the mean flux within a 50\AA ~ region centering 1500\AA, which is then subsequently converted into the absolute magnitudes. In Table \ref{tab:properties_composite}, we report the mean of the $\beta$-slope values for all targets within the six sub-samples as the representative $\beta$ of that sub-sample, and similarly, the mean of M$\rm_{UV}$ within a bin as the representative value of that bin.

\section{Discussion}
\label{sec:discussion}

\begin{figure}
    \centering
    \includegraphics[width=0.45\textwidth]{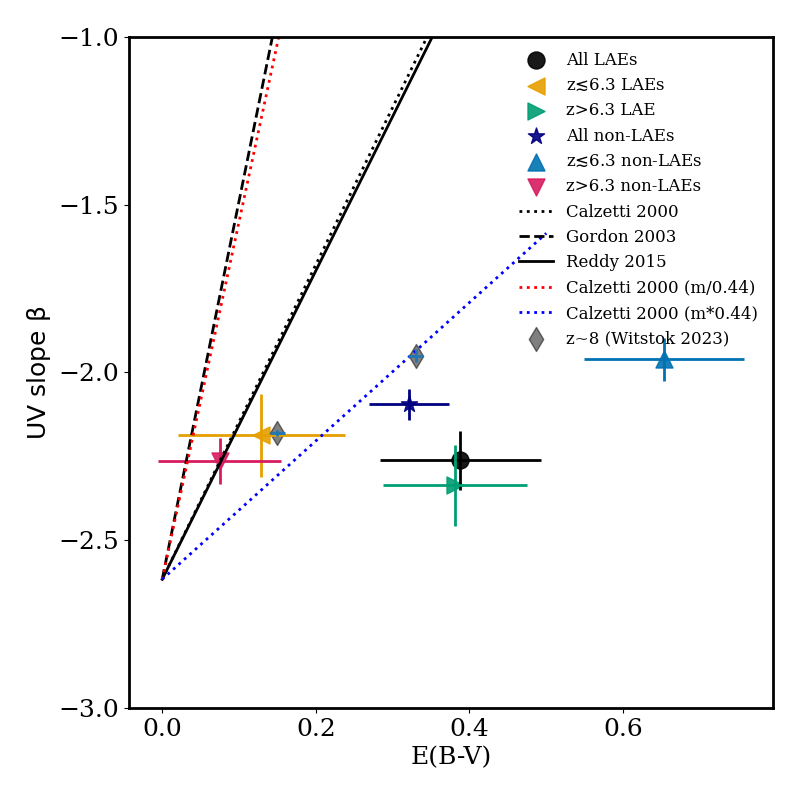}
    \caption{UV continuum slope $\beta$ versus E(B-V) for all 6 composite spectra in Section \ref{sec:composite}, along with the corresponding values of composite spectra from \citet{Witstok2023} (grey diamonds). The black solid, dotted and dashed straight lines denote the $\beta$ versus E(B-V) relations taken from \citet{Reddy2018} corresponding to the extinction curves of \citet{Reddy2015, Calzetti2000} and \citet{Gordon2003}. It is not clear from \citet{Reddy2018} whether the relation ($\beta$ = c + m*E(B-V)) corresponding to \citet{Calzetti2000} denotes the nebular or stellar E(B-V) which differs by a factor of 0.44. We account for this factor via the dotted red and blue straight lines.}
    \label{fig:ebv-beta}
\end{figure}

\subsection{Interstellar dust}
\label{sec:dust}
\indent Figure \ref{fig:ebv-beta} shows the variation of the average UV slope $\beta$ with the color-excess E(B-V) for each sub-sample. Note that the average UV slope is the mean of slopes of all spectra within a sub-sample, while the E(B-V) is directly obtained from the composite spectra. The E(B-V) values for all composite spectra, both LAEs and non-LAEs, lie within the E(B-V) range (0-1.19) found for the star-forming galaxies in the redshift range of z$\sim$2.7--6.5 by \citet{shapley2023b}. If no redshift separation is applied, LAEs show similar color excess as the non-LAEs. We find that the LAEs at z$\lesssim$6.3 show lower dust attenuation than those at z$>$6.3. This could result due to LAEs following peculiar extinction laws \citep{Atek2014} at high redshifts discussed later. On the other hand, the non-LAEs at z$\lesssim$6.3 show higher dust attenuation than those at z$>$6.3. It is possible that the results are affected by systematic uncertainties, though we do not find any direct evidence of such an effect (see Appendix \ref{app:unc}). 

\indent In Figure \ref{fig:ebv-beta}, we also over plot three different $\beta$ versus E(B-V) relations presented in \citet{Reddy2018}, and corresponding to different extinction laws for fiducial 0.14 Z$_{\odot}$ \textsc{bpass} models. These extinction curves include that of \citet{Calzetti2000}, \citet{Gordon2003} and \citet{Reddy2015}. It is not clear from \citet{Reddy2018} whether their relation corresponding to \citet{Calzetti2000} extinction curve is for nebular or stellar E(B-V), which differs by a factor of 0.44. We consider this factor by including two more straight lines (red and blue) obtained by dividing and multiplying the slope of the relation by 0.44. Note that we have estimated E(B-V) following the extinction law of \citet{Cardelli1989}. However, the E(B-V) obtained by \hg/\hb~will remain the same irrespective of extinction curves used because of similar behavior of extinction curves in the optical range. Figure \ref{fig:ebv-beta} shows that the relations from \citet{Reddy2018} do not necessarily hold for all sub-samples irrespective of their redshifts or \lya~ emission. However, the blue dotted straight line seems to roughly satisfy the values reported for composite spectra data from \citet{Witstok2023} at z$\sim$8. The lack of evident agreement between data and $\beta$-E(B-V) relations might indicate that the ISM environment of the high-z galaxies does not follow the same attenuation law as the low redshift. Another possibility is that dust is not the prime driver for determining the UV slope $\beta$ at the high redshifts \citep[see e.g.,][]{Wilkins2011}. Apart from the dust content, $\beta$ is known or speculated to be sensitive to various parameters such as metallicity, age of the stellar population, initial mass function \citep{Bouwens2010}, nebular and stellar continuum \citep{Raiter2010}, and star-formation history \citep{Schaerer2005}.

\begin{figure}
    \centering
    \includegraphics[width=0.5\textwidth]{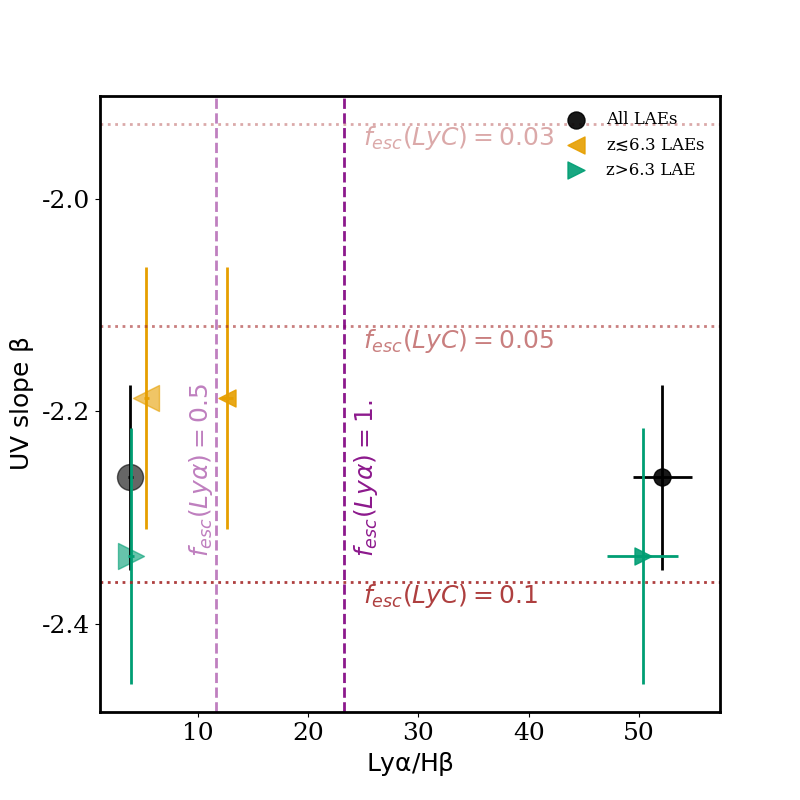}
    \caption{UV slope $\beta$ versus \lya/\hb ~for 3 composite spectra obtained from stacking the R1000 spectra of LAE sub- samples. Different samples are: (a) All LAEs with EW(\lya) $>$ 20\AA ~(black circle); (b) Targets with z$\lesssim$6.3 (orange left-pointing triangle); (c) Targets with z$>$6.3 (green right-pointing triangle). The darker legends correspond to ratios obtained by using dust-corrected line fluxes, while the lighter legends correspond to those without dust correction. The dotted horizontal lines correspond to \fescC = 3\%, 5\% and 10\% and are obtained by using  \fescC-$\beta$ relation from \citet{Chisholm2022}. The dashed vertical lines correspond to \fescLy = 0.5 and 1, and are obtained by assuming a theoretical value of \lya/\hb = 23.3. The dust corrections for the samples (a, black circle triangle) and (c, dark green triangle) are very large, which results in very large values of \lya/\hb~and consequently \fescLy ($>$ 1).}
    \label{fig:beta-LyHb}
\end{figure}

Figure \ref{fig:beta-LyHb} shows UV slope $\beta$ versus \lya/\hb~derived for three LAEs sub-samples.  We show \lya/\hb~before and after dust correction via light and dark legends, respectively. We find that dust-corrected \lya/\hb~is sometimes much larger than the theoretical value for a Case B recombination, which results in \fescLy $>$ 1. In particular, we find that \fescLy~ increases with E(B-V) for the LAE sub-samples when we compare Figures \ref{fig:ebv-beta} and \ref{fig:beta-LyHb}. The result is in agreement with previous studies and has been attributed to several physical processes or their combination at play. Firstly and as discussed earlier in the context of $\beta$ versus E(B-V) relation (Figure \ref{fig:ebv-beta}), LAEs at high redshifts may follow a peculiar dust extinction law, which is shaped by the ISM geometry of an inhomogeneous multi-phase ISM \citep{Atek2014}. In this scenario proposed by \citet{Neufeld1991}, dusty gas clouds reside within neutral small clouds in an intercloud medium with negligible absorption and scattering coefficients. Under suitable conditions, \lya~photons, being prone to resonant scattering, will bounce off from one cloud to the other without suffering any dust attenuation, while the non-resonant photons from the Balmer series (such as \ha~or \hb) pass through the neutral medium unaltered and suffer dust attenuation, thus resulting in a higher \fescLy. Such an environment will also result in an enhanced EW(\lya), i.e., line-to-continuum photons, with an increased E(B-V), which can be explored further by studying dust attenuation in individual LAEs. The enhanced \lya/\hb~has also been attributed to other arguments related to ISM geometry such as clumpy ISM resulting in a non-typical effective dust attenuation \citep{Scarlata2009, Natta1984}, and accounting for more realistic ISM configurations (other than spherical geometry) favoring \lya~transmission including ionized cones and sources' inclination angles \citep{Atek2014}.       

\indent Another possibility for large values of \lya/\hb~ could be the origin of \lya~emission. While \lya~escape fraction is generally estimated assuming the theoretical value of \lya/\ha~flux ratio in the case B recombination, this ratio can be higher in the case of collisional excitation \citep{Dijkstra2019} causing the \lya~emission in the outer region of a galaxy \citep {Ouchi2020} via cold accretion \citep[see e.g.,][]{Dijkstra2009, Goerdt2010} and possibly even outflows \citep{Daddi2021}. If the ISM temperature is higher than $\sim$2$\times$10$^4$K, the collisional excitation and de-excitation can dominate the recombination process \citep{Atek2014}. All sub-samples of LAEs, including those with \fescLy $>$ 1,  show the electron temperatures $\sim$ 2$\times$10$^4$K (see Table \ref{tab:properties_composite}).

\begin{figure}
    \centering
    \includegraphics[width=0.45\textwidth]{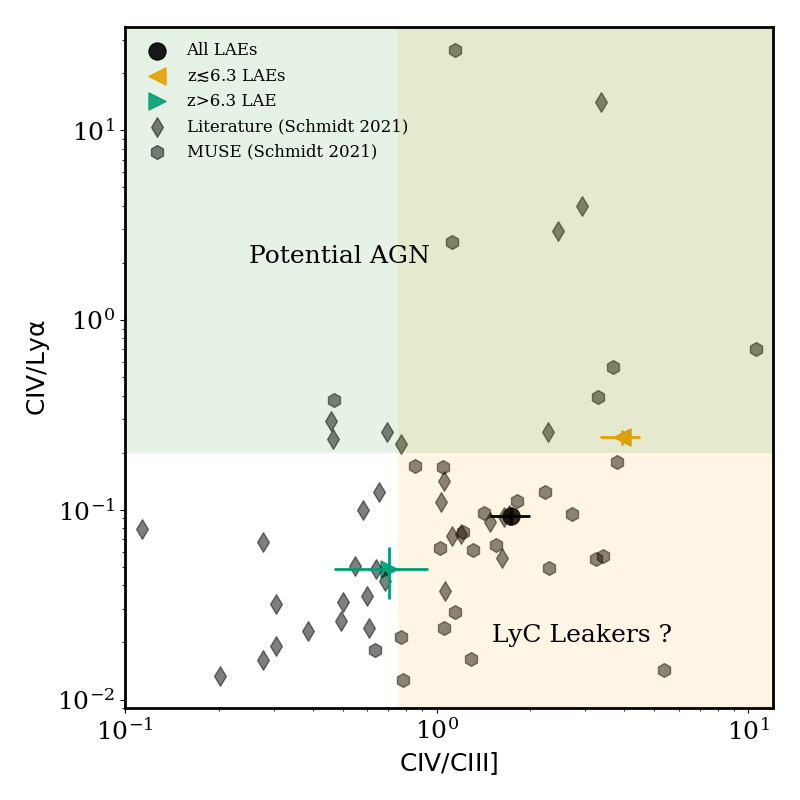}
    \caption{Variation of \civ/\lya~ with respect to \civ/\ciii~ for the LAE stacks and the literature data from \citet{Schmidt2021}. The ratios corresponding to the non-LAE sample are not shown as \lya~line and \civ~line were not detected with enough S/N. The yellow shaded region indicates \civ/\ciii$>$0.75, which is proposed by \citet{Schaerer2022} as a potential identifier of LyC leakers. The green shaded region indicates \civ/\lya$>$0.2 and indicates potential AGN.}
    \label{fig:c43_c4lya}
\end{figure}

\indent \fescLy~is expected to be connected to \fescC, because both of these quantities are affected by the geometry and the nature of gas in and around the young star-forming region. A positive correlation is found between \fescLy and \fescC in both observational \citep{Flury2022b, Begley2024} and theoretical studies \citep{Dijkstra2016, Kimm2019, Maji2022}. In Figure \ref{fig:beta-LyHb}, we also show \fescC~ corresponding to $\beta$ values from the prescriptions of \citet{Chisholm2022}. We find that \fescC~ of the three LAE samples agree with each other considering uncertainties, though it appears that LAEs at z$>$6.3 have higher \fescC~ than those at z$\lesssim$6.3. The $\beta$-derived \fescC~for the sub-sample containing all LAEs and those at z$\lesssim$6.3 are less than 10\%, and contradict the suggestion from \citet{Schaerer2022}, which would categorize both of these LAE samples as strong LyC leakers (i.e.,  \fescC~$>$0.1) owing to their high \civ/\ciii~ values ($>$0.75, see Figure \ref{fig:c43_c4lya}). We do not further comment upon \fescC~estimated here because \citet{Choustikov2023} show that $\beta$ is a good indicator of \fescC~only for metallicities $>$ 0.1 solar metallicity. \citet{Kumari2024} also find that \fescC~estimated from $\beta$ is negligible compared to that estimated from far-infrared emission lines for a local analog of reionization era galaxy, Pox 186. Similarly, \civ/\ciii ~for Pox 186 was too low to categorize this dwarf as a LyC leaker \citep{Kumari2024}. Physical measurables (other than $\beta$) are also found to correlate with \fescC, though they are also sensitive to other physical quantities. We could not estimate \fescC~ from a recent 6-parameters-based recipe from \citet{Choustikov2023} as three of these parameters (O32, R23, and E(B-V)) are determined from the composite spectra and the other two parameters ($\beta$, M$\rm_{UV}$) are the average values from galaxies in different samples, thus leading to a heterogeneous determination of parameters and hence not suitable. Moreover, the sixth parameter \hb~luminosity is not possible to determine for each of the sub-samples as \hb~ is not necessarily detected for all galaxies. Such a comparison will be presented in Bunker et al. (in prep).

\subsection{Ionization source}
\label{sec:ionization}

\begin{figure}
    \centering
    \includegraphics[width=0.45\textwidth]{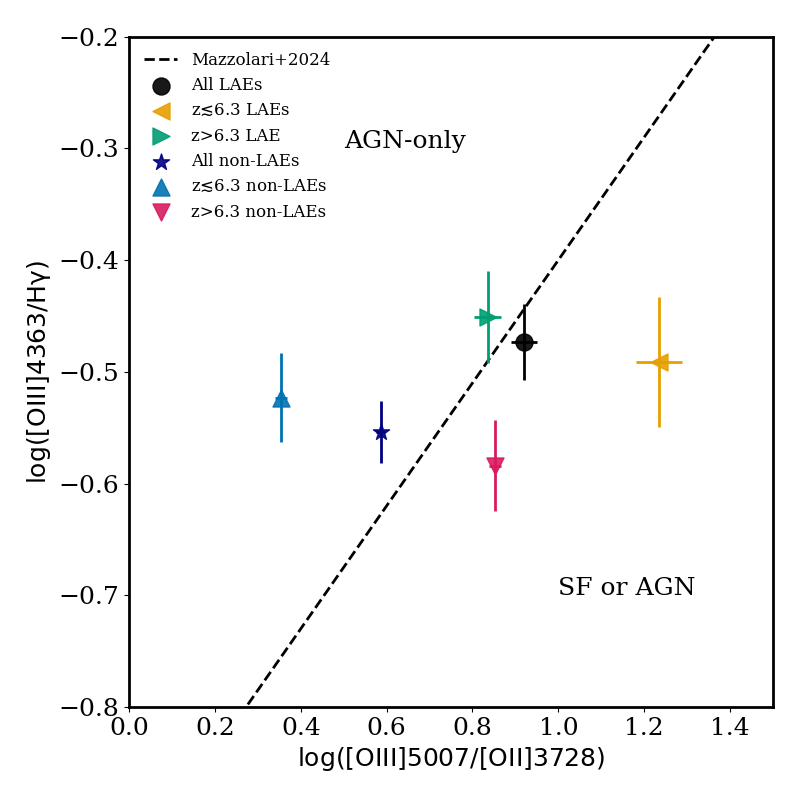}
    \caption{\oiii4363/\hg~versus \oiii5007/\oii3727,3729 ~for the six composite spectra obtained from stacking the R1000 spectra of different samples: (a) LAEs with EW(\lya) $>$ 20\AA ~(black circle); (b) LAEs with z$\lesssim$6.3 (orange left-pointing triangle); (c) LAEs with z$>$6.3 (green right-pointing triangle); (d) non-LAEs (dark blue star); (e) non-LAEs with z$\lesssim$6.3 (light-blue upward-pointing triangle) (f) non-LAEs with z$>$6.3 (magenta downward-pointing triangle). The dashed line corresponds to the demarcation line devised by \citet{Mazzolari2024}. The region above this demarcation line corresponds to an AGN-only population, while the region below this line corresponds to a population that can be either AGN or star-forming.}
    \label{fig:O3Hg_O32}
\end{figure}

\indent The composite spectra of all LAE samples (Figure \ref{fig:stacks}) show clear detection of \civ~thus manifesting that the LAEs are capable of producing very high energy photons (with energies in excess of 47.9 eV). In Figure \ref{fig:c43_c4lya}, we find that there is a large variation in both \civ/\ciii~  and \civ/\lya~line ratios. The LAEs at z$\lesssim$6.3 have \civ/\ciii~ and \civ/\lya~ about 5 times larger than the LAEs at z$>$6.3. Similarly, EW(\civ) of LAEs at z$\lesssim$6.3 is $\sim$4 times larger than that at z$>$6.3 (see Table \ref{tab:properties_composite}), and are consistent with the narrow-lined AGNs \citep{Alexandroff2013}.  \civ/\ciii~ versus \ciii/\heii~ has been explored for distinguishing the sources of ionization as star-forming, composite, and AGN \citep{Feltre2016, Hirschmann2023, Scholtz2023}. \heii1640 is detected in only 2 of the composite spectra, i.e., the one consisting of all LAEs and the other one with LAEs at z$\lesssim$6.3. Their log(\ciii/\heii) values are too low ($\sim$-0.35), thus classifying the two samples as AGN-dominated \citep{Hirschmann2023}. 

\indent Unlike UV lines, several optical lines are detected in all composite spectra, enabling us to explore the emission line ratio diagnostics diagrams involving the optical lines. However, several works have demonstrated that most of the existing optical emission line diagnostic diagrams fail to distinguish the source of ionization in high-redshift objects. The current work also confirms this (see Appendix \ref{app:fail-line-diagnostics}). In Figure \ref{fig:O3Hg_O32}, we explore \oiii4363/\hg~versus \oiii/\oii~ diagnostic diagram proposed recently by \citet{Mazzolari2024}, to distinguish the AGN-only versus SF or AGN population. This figure shows that AGN may dominate the nebular emission for a significant fraction of galaxies, and hence dominate the auroral emission in the composite spectra. 

\indent From the above analysis, it is not obvious what is the dominant feature that may preferentially select AGN. However, it seems to suggest that at high redshifts, the presence of an AGN might aid in creating ionized bubbles that help \lya~ to escape and become visible.

\begin{figure}
\centering
    \includegraphics[width=0.5\textwidth]{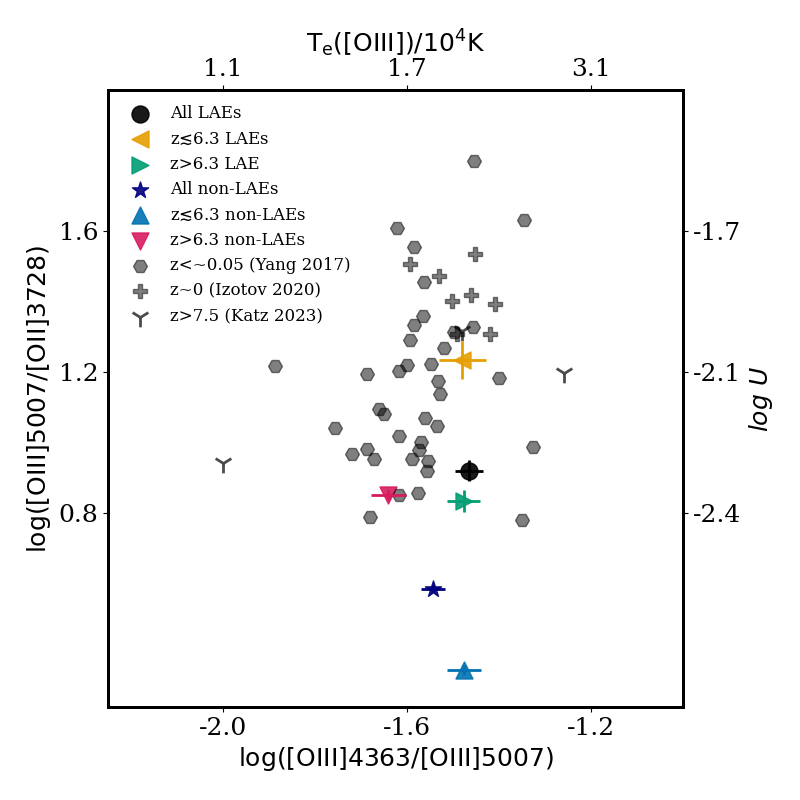}
    \caption{log(\oiii5007/\oiii4363) versus log(\oiii 5007/\oii3727,3729) for all the six composite spectra obtained from stacking the R1000 spectra of different samples. Corresponding values of Te(\oiii) and log U are also shown on the upper x- and right y-axes, respectively. Different samples are: (a) LAEs with EW(\lya) $>$ 20\AA ~(black circle); (b) LAEs with z$\lesssim$6.3 (orange left-pointing triangle); (c) LAEs with z$>$6.3 (green right-pointing triangle); (d) non-LAEs (dark blue star); (e) non-LAEs with z$\lesssim$6.3 (light-blue upward-pointing triangle) (f) non-LAEs with z$>$6.3 (magenta downward-pointing triangle). We also show the line ratios of individual local galaxies from \citet{Yang2017b, Izotov2020} (grey pluses and hexagons) and reionization era galaxies from \citet{Katz2023} (grey down triangle).}
    \label{fig:RO3-O32}
\end{figure}

\begin{figure}
\includegraphics[width=0.45\textwidth]{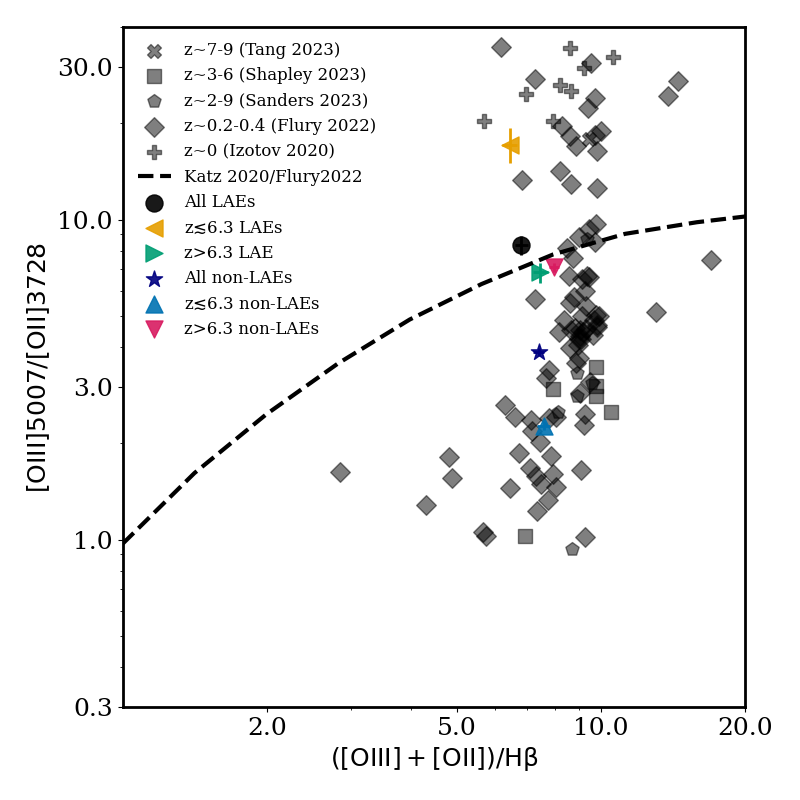}
\caption{\oiii5007/\oii3727,3729 versus R23 for the six composite spectra obtained from stacking the R1000 spectra of different samples: (a) LAEs with EW(\lya) $>$ 20\AA ~(black circle); (b) LAEs with z$\lesssim$6.3 (orange left-pointing triangle); (c) LAEs with z$>$6.3 (green right-pointing triangle); (d) non-LAEs (dark blue star); (e) non-LAEs with z$\lesssim$6.3 (light-blue upward-pointing triangle) (f) non-LAEs with z$>$6.3 (magenta downward-pointing triangle). We also overplot the line ratios obtained from composite spectra at various redshifts from the literature: z$\sim$2--9 \citep[grey pentagons]{Sanders2023}, z$\sim$3--6 \citep[grey squares]{Shapley2023} and z$\sim$7--9 \citep[grey crosses]{Tang2023}. Grey pluses indicate the line ratios for local compact star-forming galaxies with EW(\lya)=45--190\AA~ \citep{Izotov2020}, and grey diamonds indicate the LyC leaking galaxies from \citet{Flury2022}. The black dashed curve shows the locus of EoR galaxies predicted by the cosmological simulations of \citet{Katz2020} and is taken from \citet{Flury2022b}.}
\label{fig:R23_O32}
\end{figure}

\begin{figure}
\includegraphics[width=0.45\textwidth]{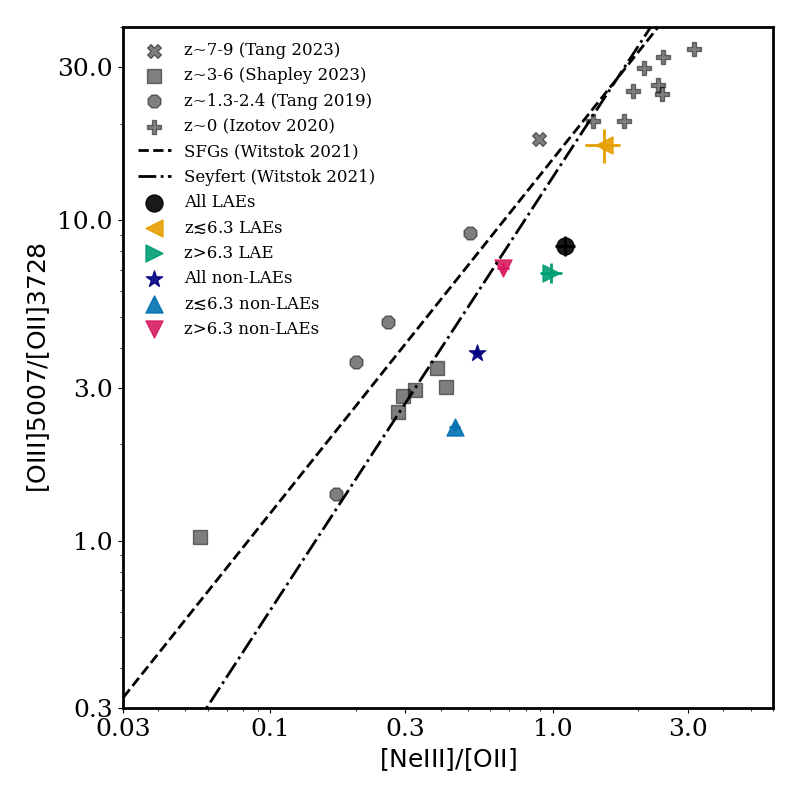}
\caption{\neiii3869/\oii3727,3729 versus \oiii5007/\oii3727,3729 for the six composite spectra obtained from stacking the R1000 spectra of different samples: (a) LAEs with EW(\lya) $>$ 20\AA ~(black circle); (b) LAEs with z$\lesssim$6.3 (orange left-pointing triangle); (c) LAEs with z$>$6.3 (green right-pointing triangle); (d) non-LAEs (dark blue star); (e) non-LAEs with z$\lesssim$6.3 (light-blue upward-pointing triangle) (f) non-LAEs with z$>$6.3 (magenta downward-pointing triangle). We also overplot the line ratios obtained from composite spectra at various redshifts from the literature: z$\sim$1--2 \citep[grey circles]{Tang2019}, z$\sim$3--6 \citep[grey squares]{Shapley2023} and z$\sim$7--9 \citep[grey crosses]{Tang2023}. Grey pluses indicate the line ratios for local compact star-forming galaxies with EW(\lya)=45--190\AA \citep{Izotov2020}. We also show the  \neiii/\oii ~versus \oiii/\oii ~relations for star-forming galaxies (black dashed line) and Seyfert (black dash-dot line) derived from the local SDSS sample by \citet{Witstok2021}.}
\label{fig:Ne3O2_O32}
\end{figure}

\begin{figure}
    \centering
    \includegraphics[width=0.45\textwidth]{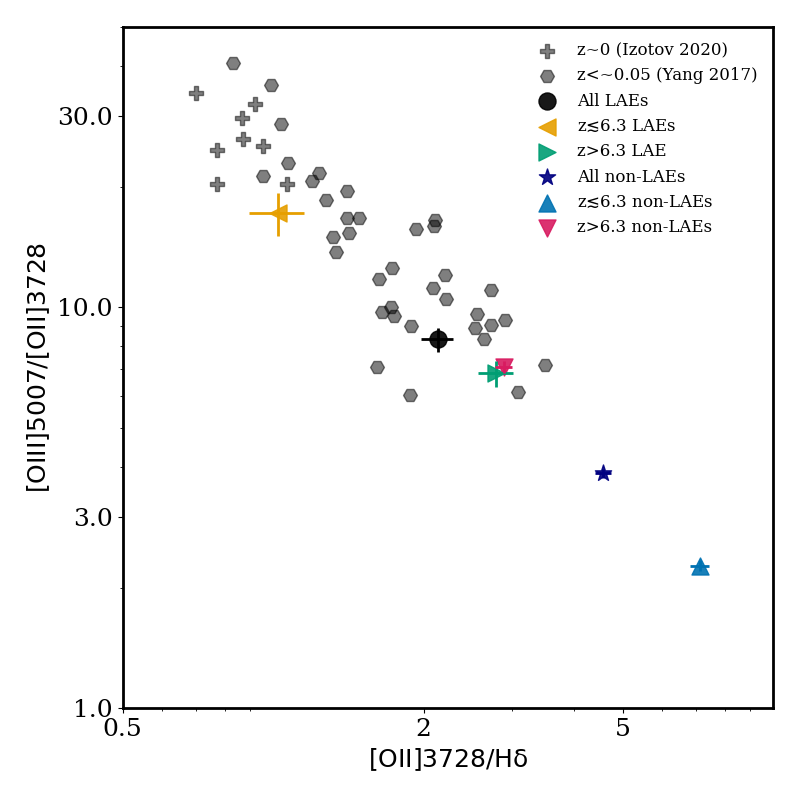}
    \caption{\oiii5007/\oii3727,3729 versus \oii3727,3729/\hd~ for the six composite spectra obtained from stacking the R1000 spectra of different samples: (a) LAEs with EW(\lya) $>$ 20\AA ~(black circle); (b) LAEs with z$\lesssim$6.3 (orange left-pointing triangle); (c) LAEs with z$>$6.3 (green right-pointing triangle); (d) non-LAEs (dark blue star); (e) non-LAEs with z$\lesssim$6.3 (light-blue upward-pointing triangle) (f) non-LAEs with z$>$6.3 (magenta downward-pointing triangle). We also overplot the corresponding line ratios of \lya~emitters at z$\sim$0 from \citet{Izotov2020}, and local analogs of \lya~emitters from \citet{Yang2017b}.}
    \label{fig:O32_O2Hd}
\end{figure}

\subsection{\lya~ versus non-\lya~ emitters}
\label{sec:lya-non-lya}

\indent LAEs are known to show large values of \oiii/\oii~ \citep[e.g.,][]{Nakajima2016, Erb2016}. Figure \ref{fig:RO3-O32} shows the variation of log(O32) (=log(\oiii/\oii)) with respect to temperature-sensitive line ratio log(RO3) (= log(\oiii4363/\oiii5007)) for all sub-samples. Since RO3 and O32 are sensitive to T$_e$(\oiii) and log(U), respectively, we also show their values on corresponding opposite axes, along with the individual galaxies' data at different redshifts from literature including \citet[][z$<0.05$]{Yang2017b}, \citet[][z$\sim$0]{Izotov2020} and \citet[][z$>$7.5]{Katz2023}. The RO3 values are relatively small for all samples, which results in larger electron temperatures compared to those found typically at lower redshifts (7000-14000K). These large electron temperatures manifest in low gas-phase metallicities of the composite spectra. The variation of gas-phase metallicity is relatively small ($\sim$0.3 dex) compared to the variation in \oiii/\oii ($\sim$ a factor of 10) across the six sub-samples. We find the same behavior in Figure \ref{fig:R23_O32}, which shows the variation of O32 versus the indirect strong-line metallicity diagnostic R23, thus confirming the results of \citet{Nakajima2016} that low metallicity is not solely responsible for extreme values of \oiii/\oii~ within LAEs. On the other hand, log(U) shows a relatively large variation ($\sim$0.7 dex), thus indicating that a high ionization parameter could lead to high \oiii/\oii.  The ionization parameter depends on several factors, such as the hardness of the ionizing spectrum \citep{Kumari2021}, ionization photon production rate, gas density, and the volume filling factor of ionizing gas \citep{Nakajima2016, Tang2019}. 

\indent The composite spectrum corresponding to all LAEs has \oiii/\oii~ about a factor of 2 larger than that containing all non-LAEs.  At z$\lesssim$6.3, \oiii/\oii~ is $\sim$8 times stronger in LAEs than in non-LAEs. At z$>$6.3, both LAEs and non-LAEs have comparable \oiii/\oii, which is likely because \lya~ emission is purely dictated by the intervening IGM at such high redshifts. This means that \oiii/\oii~ can not be used to identify the LAEs and non-LAEs at high redshifts (z$>$6.3).

\indent Like \oiii/\oii, \neiii/\oii~is also a proxy for ionization parameter. In Figure \ref{fig:Ne3O2_O32}, we show the variation of \oiii/\oii~ with respect to \neiii/\oii for all composite spectra along with the literature data. The relative behavior of \neiii/\oii~ for different sub-samples is similar to that of \oiii/\oii. However, the dynamic range of \neiii/\oii~ across different samples is much smaller than that of \oiii/\oii. For example, even at z$\lesssim$6.3, \neiii/\oii~ is only a factor $\sim$3 larger for LAEs than non-LAEs, compared to a variation of a factor $\sim$8 for \oiii/\oii. In the same figure \ref{fig:Ne3O2_O32}, we also overplot the \oiii/\oii~versus \neiii/\oii~relations for star-forming galaxies and Seyferts derived from their local (z$\sim$0) Sloan Digital Sky Survey (SDSS) samples by \citet{Witstok2021}, along with literature data at different redshifts (0$<$z$<$9). The measurements of high-redshift galaxies do not necessarily follow the empirical relations based on the measurements of local galaxies. This is likely because there are only a few data points at the higher end of \oiii/\oii~and \neiii/\oii~values in the sample considered in \citet{Witstok2021}, and they do not extend as high as the ratios considered in this work. These empirical relations should be used in high-z studies with caution.    

\indent As for \oiii/\oii, the composite spectra of all LAEs sample and all non-LAEs sample show distinct line ratios \oii/\hd~ as shown in Figure \ref{fig:O32_O2Hd}. At z$\lesssim$6.3, \oii/\hd~ for LAEs is $\sim$ 7 times weaker than for non-LAEs. The local LAEs \citep{Yang2017b, Izotov2020} show similarly stronger \oiii/\oii~and weaker \oii/\hd~ as shown by  z$\lesssim$6.3 composite LAE spectrum. We propose that in the absence \lya~ line coverage, a galaxy at z$\sim$4.8--6.3 can be classified as an LAE if it has \oiii/\oii$\gtrsim$8 and \oii/\hd$\lesssim$2 (based on the composite spectrum of all LAEs sample), though detailed modeling of ionization sources and comparison with larger dataset would be needed to test this proposition. At z$>$6.3, \oiii/\oii~ and \oii/\hd~remain comparable for both LAEs and non-LAEs; however, the degeneracy can be broken by inspecting EW(\oiii4959, 5007), which is typically high ($>$1000\AA) for LAEs than for non-LAEs (Table \ref{tab:R100}).

\indent The $\xi^{\star}_{ion}$ for LAEs samples are systematically higher than the non-LAEs (Table \ref{tab:R100}), indicating that the LAEs are galaxies with intense radiation fields. The log($\xi^{\star}_{ion}$) for LAE samples lies in the range of 25.63-25.75 (Hz erg$^{-1}$), and are in agreement with those derived by \citet{Stark2015b, Stark2017} for three LAEs.  The composite spectra corresponding to all non-LAE samples and that of z$\lesssim$6.3 non-LAEs show log($\xi^{\star}_{ion}$) closer to the canonical values used in the reionization calculations \citep[e.g.,][]{Robertson2013, Bouwens2015}, but lower than that of z$>$6.3 non-LAEs. 

\section{Summary}
\label{sec:summary}

In this work, we have studied the UV and optical properties of a sample of 253 sources at z$\sim$4.8--9.6 observed with JWST/NIRSpec with the aim of understanding the differences between the LAE and non-LAE populations at different redshifts. The main results of this study are summarized as follows: 

\begin{enumerate}
    \item We identified 42 LAEs with EW(\lya)$\gtrsim$20 and 211 non-LAEs with weak or no \lya~ emission. We further subdivided each sub-sample of LAEs and non-LAEs on the basis of the median redshift of LAEs, i.e., z$\sim$6.28.
    \item We created six composite spectra by taking the variance-weighted mean of the R1000 spectra normalized by the UV continua estimated from R100 spectra. We used these composite spectra to estimate UV and optical emission line ratios and other dependent quantities such as electron temperatures, direct and indirect metallicities, and ionization parameters.
    \item We created another set of six composite spectra by stacking the R100 spectra normalized by the continua at \oiii+\hb~ complex. We used these composite spectra to determine the EW(\hb), EW(\oiii4959, 5007) and $\xi^{\star}_{ion}$.   
    \item We estimated average UV continuum slope $\beta$ and UV magnitude M$\rm_{UV}$ by simply taking the mean of the corresponding values for each target within a sub-sample. The $\beta$ versus E(B-V) for LAE and non-LAEs do not follow the local relations, indicating that the local dust-attenuation laws are not valid for high-z environments. 
    
    \item \lya/\hb~ for two LAE samples (EW(\lya)$>$20\AA, and z$>$6.3) show values higher than the theoretical prediction for Case B recombination, which might be related to a clumpy dust geometry or the collisional excitation dominating the recombination processes. This results in \fescLy$>$1 for these LAE samples.

    \item \civ ~doublet remains undetected in the non-LAE composite spectra while they are well-detected in the LAE samples. \civ/\lya~and \civ/\ciii~for LAE population at z$\lesssim$6.3 is $\sim$a factor of 5 larger than that for LAE population at z$>$6.3. \civ/\ciii~line ratios and UV slope $\beta$ for a given sub-sample of LAEs give contradictory information about the escape fraction of LyC.

    \item Both UV and optical line ratios suggest that AGN might dominate all sub-samples of galaxies under study irrespective of \lya~ emission and redshift. However, it is not clear how these AGNs are selected within each sub-sample. The hint of AGN in each sub-sample also suggests that they might be responsible for creating ionized bubbles and hence \lya~ escape. 

    \item LAEs have much stronger \oiii/\oii and much lower \oii/\hd~ compared to the non-LAEs at z$\lesssim$6.3. However, both of these line ratios are comparable for LAEs and non-LAEs at z$>$6.3, where a high value of EW(\oiii4959, 5007) ($>$1000\AA) is useful in breaking the degeneracy and distinguishing LAEs from non-LAEs.    

    \item LAEs have harder ionizing radiation field as revealed by \civ~ detection, high \oiii/\oii~ and large $\xi_{ion}^{\star}$ compared to the non-LAEs.
    
\end{enumerate}

\indent In summary, the present analysis allowed us to identify criteria to distinguish between the LAE and non-LAE populations in general, as well as at different redshifts. It also raised questions about the applicability of the existing dust attenuation laws at high redshifts, the use of typical emission line ratio diagnostics in identifying the ionization mechanisms within distant galaxies, and the likely role of AGN in creating ionized bubbles in the reionization era. Further detailed studies of LAEs and non-LAEs involving deeper spectroscopy ensuring the detection of continuum and \heii~ features, combined with modeling and simulations, are required to address these questions and better understand the populations of LAEs and non-LAEs across different redshifts and particularly their role in reionizing the Universe.

\begin{acknowledgments}
We thank the JWST Instrument Development Teams and the instrument teams at the European Space Agency and the Space Telescope Science Institute. RS acknowledges support from an STFC Ernest Rutherford Fellowship (ST/S004831/1). 
JW, FDE, RM \& CS acknowledge support by the Science and Technology Facilities Council (STFC), by the ERC through Advanced Grant 695671 “QUENCH”, and by the UKRI Frontier Research grant RISEandFALL.  RM also acknowledges funding from a research professorship from the Royal Society. AJB, AJC, JC, AS \& GCJ acknowledge funding from the "FirstGalaxies" Advanced Grant from the European Research Council (ERC) under the European Union’s Horizon 2020 research and innovation programme (Grant agreement No. 789056). SC acknowledges support by European Union’s HE ERC Starting Grant No. 101040227 - WINGS. ECL acknowledges the support of an STFC Webb Fellowship (ST/W001438/1).  DJE is supported as a Simons Investigator and by JWST/NIRCam contract to the University of Arizona, NAS5-02015. BER acknowledges support from the NIRCam Science Team contract to the University of Arizona, NAS5-02015, and JWST Program 3215. The research of CCW is supported by NOIRLab, which is managed by the Association of Universities for Research in Astronomy (AURA) under a cooperative agreement with the National Science Foundation. CAW acknowledges the support from the JWST/NIRCam contract to the University of Arizona NAS5-02015. This research is supported in part by the Australian Research Council Centre of Excellence for All Sky Astrophysics in 3 Dimensions (ASTRO 3D), through project number CE170100013. The authors acknowledge the use of the lux supercomputer at UC Santa Cruz, funded by NSF MRI grant AST 1828315.
\end{acknowledgments}

\vspace{5mm}
\facilities{JWST(NIRSpec/MSA, NIRCam)}

\software{astropy \citep{2013A&A...558A..33A,2018AJ....156..123A}, matplotlib \citep{Hunter2007}, numpy \citep{Harris2020}, lmfit \citep{Newville2014}, topcat \citep{Taylor2005}, WebPlotDigitizer \citep{Rohatgi2022}
          }


\appendix 

\section{R100 Composite Spectra \& Properties}
\label{app:R100_composite}

\indent For creating the R100 composite spectra, we de-redshift and resample all R100 spectra on a common wavelength grid, normalize them by the continuum level at the \oiii+\hb~ complex, and take the mean weighted by the variance in the spectra. We follow the above procedure for all six sub-samples described in Section \ref{sec:sample}. However, the continua were not detected in some of the galaxies. So, the number of galaxies within R100 composite spectra is slightly different than that in the R1000 composite spectra but still statistically significant within each bin (compare N in Tables \ref{tab:properties_composite} and \ref{tab:R100}). 

\indent We use the R100 composite spectra obtained by normalizing the spectra by the continua at \oiii+\hb~ to estimate the EW(\oiii+\hb). We integrate the flux under the \hb ~line and that under \oiii~doublet to estimate the corresponding EWs. The total fluxes from these continua-normalized composite spectra correspond to the EWs. 

\indent We estimate $\xi^{\star}_{ion}$ for each composite R100 spectra using EW(\oiii4959,5007) in the following relation from \citet{Chevallard2018}: $\rm log(\xi^{\star}_{ion}) = 22.12 + 1.15 \times log(EW(\oiii4959, 5007))$. 

\indent Table \ref{tab:R100} summarizes the properties of the R100 composite spectra estimated in this section.

\begin{table*}
\centering
\caption{Properties of the R100 composite spectra from different samples.}
\label{tab:R100}
\begin{tabular}{lcccccc}
\hline
\hline
& \multicolumn{3}{c}{LAEs} & \multicolumn{3}{c}{non-LAEs}\\
Properties & All & z$\lesssim$6.3 & z$>$6.3 & All & z$\lesssim$6.3 & z$>$6.3 \\
\hline
N & 38 & 18 & 20 & 179 & 104 & 75 \\
EW(\oiii4959, 5007) (\AA) &$1266\pm2$ &$1430\pm3$ &$1119\pm3$ &$597\pm1$ & $571\pm1$&$787\pm2.$ \\
EW(\hb) (\AA) & $282\pm1$&$307\pm2$ &$239\pm$2 & $160\pm$1& $157\pm1$&$179\pm2$ \\
log($\xi^{\star}_{ion}$ /Hz erg$^{-1}$) & 25.68 & 25.75 & 25.63 & 25.31 & 25.29 & 25.45 \\
\hline
\end{tabular}
\end{table*}

\begin{figure}
    \centering
    \includegraphics[width=1.\textwidth]{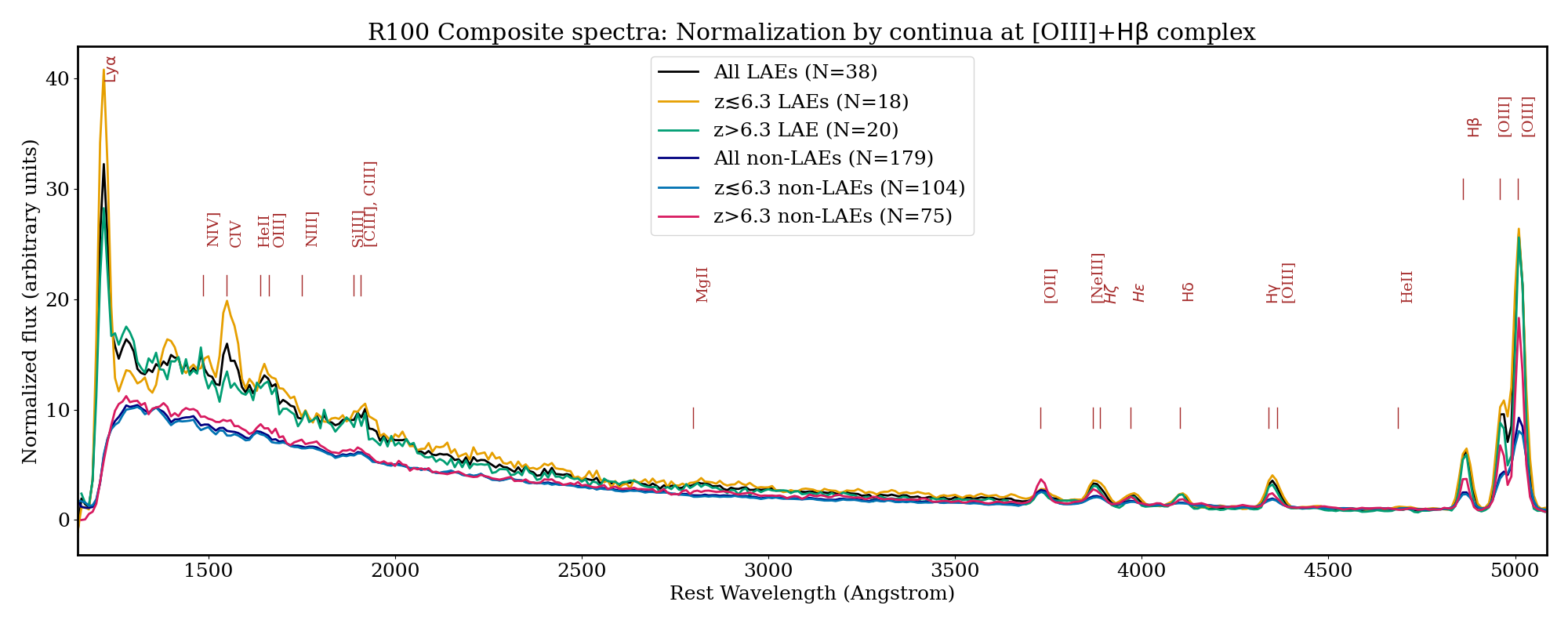}
    \caption{R100 composite spectra obtained by normalizing individual R100 spectra with respect to the continua at \oiii+\hb~ complex. Different colors correspond to the six sub-samples of targets described in Section: (a) 38 LAEs with EW(\lya)$>$20\AA (black) (b) 18 LAEs lying below the median redshift z$\lesssim$6.3 (orange) (c) 20 LAEs with z$>$6.3 (green) (d) 179 non-LAEs (dark blue) (e) 104 non-LAEs with  z$\lesssim$6.3 (light blue) (f) 75 non-LAEs with z$>$6.3 (magenta). The location of the typical UV and optical emission lines is marked; however, we note that not all marked lines are detected. Unlike Figure \ref{fig:stacks}, composite spectra here are not offset as the spectral features of interest are clearly visible.}
    \label{fig:R100}
\end{figure}

\section{Uncertainties analysis}
\label{app:unc}
\citet{Deugenio2024} report a relative flux uncertainty of 15\% between the prism and gratings data. To investigate its effect on E(B-V) values of the composite spectra, we also create composite spectra weighted by the variance from the uncertainties accounting for the uncertainties in the R1000 spectra, dispersion in R100 spectra in the 1500\AA~ wavelength window and a 15\% relative uncertainty between R1000 and R100 spectra. We then fit single Gaussian to the \hg~ and \hb~ emission lines by taking the above-estimated uncertainties into account and, thus, we estimate E(B-V) from these line fluxes. We compare these E(B-V) with those estimated from composite spectra where the systematic flux uncertainty of 15\% is not taken into account, and we give equal weight to every data point while fitting the emission lines. The E(B-V) obtained from the two methods agree with each other within uncertainties.

\section{Ambiguous ionization sources}
\label{app:fail-line-diagnostics}

\begin{figure*}
\centering
    \includegraphics[width=0.45\textwidth]{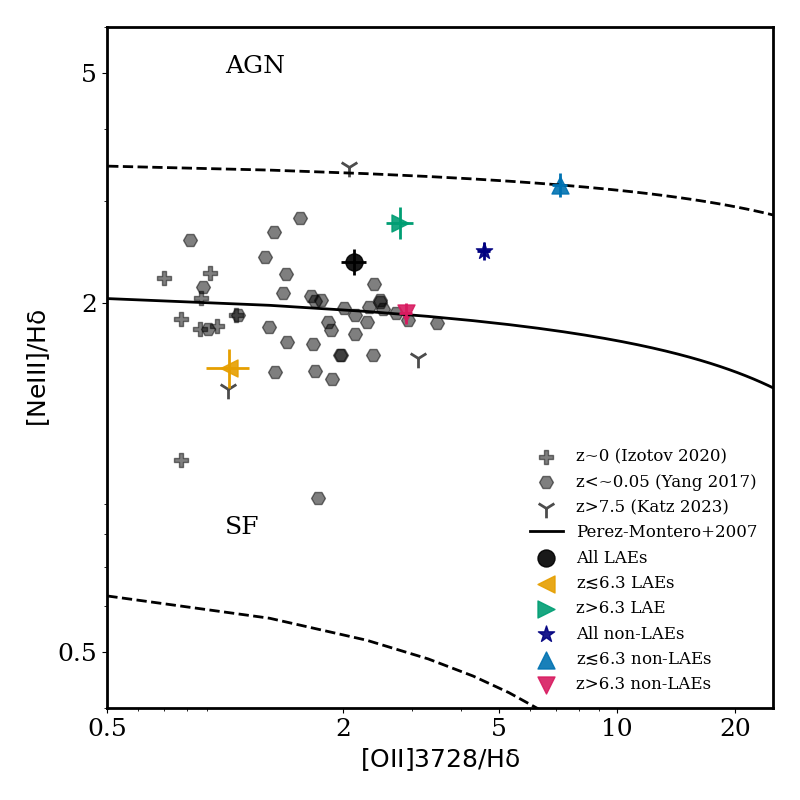}
   \includegraphics[width=0.45\textwidth]{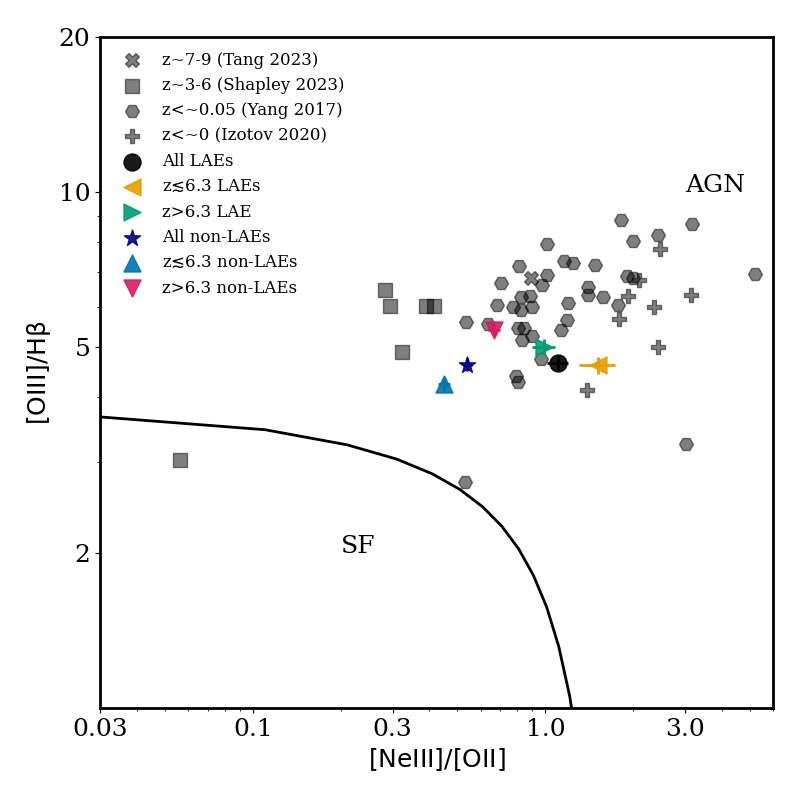} 
\caption{Emission line ratio diagnostic diagrams: \oii3727,3729/\hd~ versus \neiii3869/\hd~ (left-hand panel) and \neiii3869/\oii3727,3729 versus \oiii5007/\hb ~(right-hand panel) for all composite spectra obtained from stacking the R1000 spectra of the six sub- samples: (a) LAEs with EW(\lya) $>$ 20\AA ~(black circle); (b) LAEs with z$\lesssim$6.3 (orange left-pointing triangle); (c) LAEs with z$>$6.3 (green right-pointing triangle); (d) non-LAEs (dark blue star); (e) non-LAEs with z$\lesssim$6.3 (light-blue upward-pointing triangle) (f) non-LAEs with z$>$6.3 (magenta downward-pointing triangle). On both panels, grey pluses indicate the line ratios for local compact star-forming galaxies with EW(\lya)=45--190\AA~ \citep{Izotov2020}, and grey pentagons indicate the blueberries from \citet{Yang2017b}. On the left-hand panel, we also include ratios corresponding to  z$>7.5$ galaxies from \citet{Katz2023}. The black curve denotes the SF-AGN separator from \citet{Perez-Montero2007}. All the plotted data from this work and literature lie within the 0.15 dex uncertainty limit (dashed curves) on the SF-AGN separator. On the left panel, we also overplot the line ratios obtained from composite spectra at various redshifts from the literature: z$\sim$3--6 \citep[grey squares]{Shapley2023} and z$\sim$7--9 \citep[grey crosses]{Tang2023}. The black curve on the right panel denotes the SF-AGN separator from \citet{Backhaus2022}, and systematically classifies SF galaxies as AGNs.}
\label{fig:bpt}
\end{figure*}

In Figure \ref{fig:bpt}, we use two emission-line diagnostic diagrams: \oii/\hd~versus \neiii/\hd~(left-hand panel) and \neiii/\oii~versus\oiii/\hb ~(right-hand panel). The former was proposed by \citet{Perez-Montero2007} and provided an empirical SF-AGN separator (black curve on the left-hand panel in Figure \ref{fig:bpt}). The source of ionization for the LAE and non-LAE samples can not be distinguished as their line ratios lie within the uncertainty range of 0.15 dex along the SF-AGN separator. However, they occupy the same parameter space as the local star-forming galaxies presented in \citet{Izotov2020} and \citet{Yang2017b}. Similarly, the SF-AGN separator curve devised by \citet{Backhaus2022} classifies the LAE and non-LAE samples as AGNs, which occupy the same parameter space as those occupied by star-forming galaxies at high-redshifts \citep{Tang2019, Shapley2023} and their local analogs \citep{Izotov2020, Yang2017b}. Our result agrees with that of \citet{Cleri2023}, who also report that the \oiii/\hb-\neiii/\oii~is not a good diagnostic of the ionizing source. 

\bibliography{sample631}{}
\bibliographystyle{aasjournal}

\end{document}